\documentclass[10pt]{article}
\usepackage{amssymb,amsmath}
\usepackage{amsfonts}
\usepackage{eufrak}
\usepackage{graphicx}

\begin{document}

\begin{center}
\null\vspace{2cm}
{\large {\bf Longitudinal Wave Propagation in Relativistic Two-fluid Plasmas around Reissner-Nordstr\"om Black Hole}}\\
\vspace{2cm}
M. Atiqur Rahman\footnote{E-mail: $atirubd@yahoo.com$}\\
{\it Department of Applied Mathematics,\\ Rajshahi University,\\
Rajshahi - 6205, Bangladesh}
\end{center}
\vspace{3cm}
\centerline{\bf Abstract}
\baselineskip=18pt
\bigskip

The 3+1 spacetime formulation of general relativity is used to investigate the transverse waves propagating in a plasma influenced by the gravitational field of Reissner-Nordstr\"om black hole, as explained in an earlier paper, to take account of relativistic effects due to the event horizon. Here, a local approximation is used to investigate the one-dimensional radial propagation of longitudinal waves. We derive the dispersion relation for these waves and solve it numerically for the wave number $k$.
\vspace{0.5cm}\\
{\it PACS number(s)}: 95.30.Sf, 95.30.Qd, 97.60.Lf\\
{\it Keywords}: Longitudinal waves, Two-fluid plasma, Charged black hole

\vfill

\newpage
\section{Introduction}\label{sec1}

In recent years there has been renewed interest in investigating plasmas in the black hole environment. The Coulomb potential of charge particles due to coupling is much stronger than the gravitational potential and are often neglected in the Newtonian approximation. But the mean gravitational field for certain astronomical objects like galactic nuclei or black holes may be strong and the observation of magnetic fields indicates that a combination of general relativity and plasma physics at least on the level of a fluid description is appropriate. A successful study of the waves and emissions from plasmas falling into a black hole will be of great value in aiding the observational identification of black hole candidates. For this reason, plasma physics in the vicinity of a black hole has become a subject of great interest in astrophysics. In the immediate neighborhood of a black hole general relativity applies. It is therefore of interest to formulate plasma physics problems in the context of general relativity.

In the preceding paper \cite {one}, a local approximation has been used to investigate the transverse wave propagation in the relativistic two-fluid plasmas surrounding the Reissner-Nordstr\"om (RN) black hole event horizon. The present paper is concerned with the investigation of longitudinal waves propagation in this environment. A general introduction to plasma physics in the presence of the gravitational field of black hole has been presented in the preceding paper I, based on the 3+1 formulation of Thorne, Price, and MacDonald (TPM) \cite{two,three,four}, given detailed in the membrane paradigm \cite{five}, which provides a foundation for formulation of a general relativistic set of plasma physics equations in the strong gravitational field of black hole. Actually the $3+1$ approach was originally developed in 1962 by Arnowitt, Deser, and Misner \cite{six} to study the quantization of the gravitational field. Since then, their formulation has most been applied in studying numerical relativity \cite{seven}.

Exploiting the 3+1 formalism, a lot of researches have been carried out. The study of plasma wave in the presence of strong gravitational fields using the $3+1$ approach is still in its early stages. Zhang \cite{eight,nine} has considered the care of ideal magneto hydrodynamics waves near a Kerr black hole, accreting for the effects of the holes angular momentum but ignoring the effects due to the black hole horizon. Holcomb and Tajima \cite{ten}, Holcomb \cite{eleven}, and Dettmann et. al. \cite{twelve} have considered some properties of wave propagation in a Friedmann universe. Daniel and Tajima \cite{thirteen} have studied the physics of high frequency electromagnetic waves in a strong Schwarzschild plasma.

The oscillations of a perfect relativistic plasma, as well as of nonrelativistic plasma, can be divided into longitudinal (electrostatic) and electromagnetic (transverse) oscillations. Buzzi, Hines, and Treumann (BHT) \cite{fourteen,fifteen} described a general relativistic version of two-fluid formulation of plasma physics and developed a linearized treatment of plasma waves in analogy with the special relativistic formulation of Sakai and Kawata (SK) \cite{sixteen}, to investigate the nature of the transverse and longitudinal waves near the horizon of the Schwarzschild black hole \cite{fourteen,fifteen}. In this paper we apply linearized two-fluid equations of BHT to investigate longitudinal waves propagating in the plasma close to the event horizon of Reissner-Nordtsr\"om (RN) black hole. The RN spacetime is the Schwarzschild spacetime generalization with charge $q$ and mass $M$. It is the unique, asymptotically flat, spherically symmetric solution of the Einstein-Maxwell equations that may be analytically extended to an electrovacuum solution representing a black hole for $0<|q|<M$. The extrime RN black hole (i.e., when $q=M$) is distinguished by its coldness (vanishing Hawking temperature) and its supersymmetry. It occupies a special position among the solutions to Einstein or Einstein-Maxwell equations because of its complete stability with respect to both classical and quantum process permitting its interpretation as a soliton \cite{seventeen,eighteen}. The extremal RN space is also special in admitting supersymmetry in the contest of $N=2$ supergravity \cite{nineteen,twenty,twenty one,twenty two,twenty three}. Recently, the quantization of longitudinal electric waves in plasma are studied in \cite{twenty four} and new longitudinal waves are found to exist for these types of quantum plasmas \cite{twenty five}. So, our study on longitudinal electric waves around a charged black hole is important and meaningful.

This paper is organized as follows. In section 2, we summarized the nonlinear two-fluid equations expressing continuity and conservation of energy and momentum in Reissner-Nordstr\"om black hole spacetime. For zero gravitational field these equations reduce to the corresponding special relativistic expressions. In section 3, the two-fluid equations for longitudinal waves have been derived by considering one-dimensional wave propagation in the radial $z$ (Rindler coordinate system) direction. We linearize these equations for wave propagation in section 4 by giving a small perturbation to fields and fluid parameters. In section 5, we express the derivatives of the unperturbed quantities with respect to $z$ and used the local or mean-field approximation to obtain numerical solutions for the wave dispersion relations. We describe the dispersion relation for the longitudinal waves in section 6 and give the numerical procedure for determining the roots of the dispersion relation. In section 7, we present the numerical solutions for the wave number $k$. In section 8, we present the numerical solutions for the wave number $k$. Finally, in section 9, we present our remarks. We use units $G=c=k_B=1$.

\section{Two-fluid Equations in RN Spacetime}\label{sec2}

As mentioned in paper \cite {one}, the two-fluid plasma considered here are either electron-positron or electron-ion. The equation of continuity and Maxwell\rq s equations for each of the fluid of species $s$ with velocity ${\bf v}_s$, number density $n_s$, and the relativistic Lorentz factor $ \gamma _s$ in Reissner-Nordstr\"om spacetime are given by
\begin{equation}
\frac{\partial }{\partial t}(\gamma _sn_s)+\nabla \cdot (\alpha \gamma _sn_s{\bf v}_s)=0\label{eq1},
\end{equation}
and
\begin{eqnarray}
\nabla \cdot {\bf B}&=&0,\label{eq2}\\
\nabla \cdot {\bf E}&=&4\pi \sigma ,\label{eq3}\\
\frac{\partial {\bf B}}{\partial t}&=&-\nabla \times (\alpha {\bf E}),\label{eq4}\\
\frac{\partial {\bf E}}{\partial t}&=&\nabla \times (\alpha {\bf B})-4\pi \alpha {\bf J},\label{eq5}
\end{eqnarray}
with the charge $q_s$ and the charge and current densities are defined for each fluid species by
\begin{equation}
\sigma =\sum_s\gamma _sq_sn_s,\hspace{1.2cm}{\bf J}=\sum_s\gamma _sq_sn_s{\bf v}_s\label{eq6},
\end{equation}
where $s$ is $1$ for electrons and $2$ for positrons (or ions). The presence of lapse function $\alpha$  defined by
\begin{equation}
\alpha (r)\equiv \frac{d\tau }{dt}=\frac{1}{r}(r-r_+)^{\frac{1}{2}}(r-r_-)^{\frac{1}{2}}\label{eq7}.
\end{equation}
in all the equations signifies the general relativistic effect around a Reissner-Nordstr\"om black hole. Here, $r_\pm=M\pm \sqrt{M^2-q^2}$. All quantities are measured by a local fiducial observer (FIDO) as discussed in paper \cite {one}. The spacetime metric of our coordinates is
\begin{eqnarray}
ds^2=-\alpha^2dt^2+\frac{1}{\alpha^2}dr^2+r^2(d\theta ^2+{\rm sin}^2\theta d\varphi ^2).\label{eq8}
\end{eqnarray}
The gravitational acceleration felt by a FIDO is defined again as
\begin{equation}
{\bf a}=-\nabla {\rm ln}\alpha =-\frac{1}{\alpha }\left(\frac{M}{r^2}-\frac{q^2}{r^3}\right){\bf e}_{\hat r}\label{eq9},
\end{equation}
The energy and momentum conservation equations derived from Maxwell\rq s Eqs. (\ref{eq4}) and (\ref{eq5}) coupling with each single perfect fluid of species $s$ to the electromagnetic field as given in paper \cite {one} are
\begin{eqnarray}
\frac{1}{\alpha }\frac{\partial }{\partial t}P_s-\frac{1}{\alpha }\frac{\partial }{\partial t}[\gamma _s^2(\varepsilon _s+P_s)]-\nabla \cdot [\gamma _s^2(\varepsilon _s+P_s){\bf v}_s]+\gamma _sq_sn_s{\bf E}\cdot {\bf v}_s+2\gamma _s^2(\varepsilon _s+P_s){\bf a}\cdot {\bf v}_s=0,\label{eq10}
\end{eqnarray}
and
\begin{eqnarray}
&&\gamma _s^2(\varepsilon _s+P_s)\left(\frac{1}{\alpha }\frac{\partial }{\partial t}+{\bf v}_s\cdot \nabla \right){\bf v}_s+\nabla P_s-\gamma _sq_sn_s({\bf E}+{\bf v}_s\times {\bf B})\nonumber\\
&&+{\bf v}_s\left(\gamma _sq_sn_s{\bf E}\cdot {\bf v}_s+\frac{1}{\alpha }\frac{\partial }{\partial t}P_s\right)+\gamma _s^2(\varepsilon _s+P_s)[{\bf v}_s({\bf v}_s\cdot {\bf a})-{\bf a}]=0\label{eq11}.
\end{eqnarray}
Here, $\varepsilon _s$ is the internal energy density and $P_s$ is the fluid pressure. If, now, one sets $\alpha =1$ so that the acceleration goes to zero,these equations reduce to the corresponding special relativistic equations as given by SK \cite{sixteen}. Although these equations are valid in a FIDO frame, the transformation from the FIDO frame to the comoving (fluid) frame involves a boost velocity, which is simply the freefall velocity onto the black hole, given by
\begin{equation}
v_{\rm ff}=(1-\alpha ^2)^{\frac{1}{2}}\label{eq12}.
\end{equation}
Then the relativistic Lorentz factor $\gamma _{\rm boost}\equiv (1-v_{\rm ff}^2)^{-1/2}=1/\alpha $.

The Rindler coordinate system, in which space is locally Cartesian, provides a good approximation to the Reissner-Nordstr\"om metric near the event horizon in the form
\begin{equation}
ds^2=-\alpha^2dt^2+dx^2+dy^2+dz^2,\label{eq13}
\end{equation}
where
\begin{equation}
x=r_+\left(\theta -\frac{\pi }{2}\right),\hspace{1cm}y=r_+\varphi ,\hspace{1cm}z=2r_+\alpha ^2\label{eq14}.
\end{equation}
The standard lapse function in Rindler coordinates becomes $\alpha =z/2r_+$, where $r_+$ is the event horizon of the black hole.

\section{Longitudinal Two-fluid Equations}\label{sec3}
In this section, only a summary of the two-fluid equations relevant for the longitudinal waves will be restated. For longitudinal waves it is more convenient to work from a combination of the equation of continuity (\ref{eq1}), Poisson equation (\ref{eq3}), and the conservation of momentum equation (\ref{eq11}). We can separated the longitudinal part of these equations by considering one-dimensional wave propagating in the radial $z$ direction. Assuming $v_{sx}$, $v_{sy}$ and $v_{sz}=u_s$ as the velocity components along $x$, $y$ and $z$ direction, the radial component $e_{\hat z}$ of the equation of continuity (\ref{eq1}) and the Poisson equation (\ref{eq3}) can be written as
\begin{equation}
\frac{\partial }{\partial t}(\gamma _sn_s)+\frac{\partial }{\partial z}(\alpha \gamma _sn_su_s)=0,\label{eq15}
\end{equation}
and
\begin{equation}
\frac{\partial E_z}{\partial z}=4\pi (q_1n_1\gamma _1+q_2n_2\gamma _2).\label{eq16}
\end{equation}
The two-fluid equations for longitudinal waves can be separated by considering a new complex transverse and longitudinal fields and velocities by introducing the following complex variables
\begin{eqnarray}
v_s(z,t)&=& v_{sx}(z,t)+{\rm i}v_{sy}(z,t),\quad v_{sz}(z,t)=u_s(z,t), \nonumber\\
B(z,t)&=&B_x(z,t)+{\rm i}B_y(z,t),\quad E(z,t)=E_x(z,t)+{\rm i}E_y(z,t)\label{eq17}.
\end{eqnarray}
These give
\begin{eqnarray}
v_{sx}B_y-v_{sy}B_x=\frac{\rm i}{2}(v_sB^\ast -v_s^\ast B),\nonumber\\
v_{sx}E_y-v_{sy}E_x=\frac{\rm i}{2}(v_sE^\ast -v_s^\ast E)\label{eq18},
\end{eqnarray}
where the $\ast $ denotes the complex conjugate.
Using the above complex variables, the longitudinal part of the momentum conservation equation (\ref{eq11}) can be written as
\begin{eqnarray}
\rho _s\frac{Du_s}{D\tau }=q_sn_s\gamma _s\left(E_z+\frac{{\rm i}}{2}\left(v_sB^{\ast}-v^{\ast}_s B\right)\right)+(1-u^2_s)\rho _sa\nonumber\\
-u_s\left(q_sn_s\gamma _s{\bf E}\cdot {\bf v}_s+\frac{1}{\alpha }\frac{\partial P_s}{\partial t}\right)-\frac{\partial P_s}{\partial z},\label{eq19}
\end{eqnarray}
where ${\bf E}\cdot {\bf v}_s=\frac{1}{2}(Ev_s^\ast +E^\ast v_s)+E_zu_s$ and $\rho _s$ is the total energy density defined by
\begin{equation}
\rho _s=\gamma _s^2(\varepsilon _s+P_s)=\gamma _s^2(m_sn_s+\Gamma _gP_s)\label{eq20}
\end{equation}
with $\Gamma _g=\gamma _g/(\gamma _g-1)$. The gas constant $\gamma _g$ take the value  $4/3$ for $T\rightarrow \infty $ and $5/3$ for $T\rightarrow 0$. The ion temperature profile is closely adiabatic and it approaches $10^{12}\,K$ near the horizon \cite{twenty six}. Far from the (event) horizon electron (positron) temperatures are essentially equal to the ion temperatures, but closer to the horizon the electrons are progressively cooled to about $10^8-10^9\,K$ by mechanisms like multiple Compton scattering and synchrotron radiation. The equation of state can be expressed by using the conservation of entropy as
\begin{equation}
\frac{D}{D\tau }\left(\frac{P_s}{n_s^{\gamma _g}}\right)=0.\label{eq21}
\end{equation}
Where the time derivative in Eqs. (\ref{eq16}) and (\ref{eq18}) is defined as
\begin{equation}
\frac{D}{D\tau }\equiv \left(\frac{1}{\alpha }\frac{\partial }{\partial t}+{u_s}\frac{\partial}{\partial z} \right).\label{eq22}
\end{equation}

\section{Linearization}\label{sec4}
We linearize the above set of longitudinal two-fluid equations by considering a small perturbation. We introduce the quantities
\begin{eqnarray}
u_s(z,t)&=&u_{0s}(z)+\delta u_s(z,t),\hspace{.2cm}v_s(z,t)=\delta v_s(z,t),\nonumber\\
n_s(z,t)&=&n_{0s}(z)+\delta n_s(z,t),\hspace{.2cm}P_s(z,t)=P_{0s}(z)+\delta P_s(z,t),\nonumber\\
\rho _s(z,t)&=&\rho _{0s}(z)+\delta \rho _s(z,t),\hspace{.2cm}{\bf E}(z,t)=\delta {\bf E}(z,t),\nonumber\\
{\bf B}_z(z,t)&=&{\bf B}_0(z)+\delta {\bf B}_z(z,t),\hspace{.2cm}{\bf B}(z,t)=\delta {\bf B}(z,t).\label{eq23}
\end{eqnarray}
Here, magnetic field has been chosen to lie along the radial ${\bf e}_{\hat z}$ direction. The relativistic Lorentz factor is also linearized such that $\gamma _s=\gamma _{0s}+\delta \gamma _s,$  where
\begin{eqnarray}
\gamma _{0s}=\left(1-{\bf u}_{0s}^2\right)^{-\frac{1}{2}},\quad \delta \gamma _s=\gamma _{0s}^3{\bf u}_{0s}\cdot \delta {\bf u}_s\label{eq24}.
\end{eqnarray}
Neglecting the product of perturbation terms the conservation of entropy, Eq. (\ref{eq21}) is linearized to
\begin{equation}
\delta P_s=\frac{\gamma _gP_{0s}}{n_{0s}}\delta n_s.\label{eq25}
\end{equation}
Also the total energy density Eq. (\ref{eq20}) is linearized to
\begin{equation}
\delta \rho _s=\frac{\rho _{0s}}{n_{0s}}\left(1+\frac{\gamma _{0s}^2\gamma _gP_{0s}}{\rho _{0s}}\right)\delta n_s+2u_{0s}\gamma _{0s}^2\rho _{0s}\delta u_s,\label{eq26}
\end{equation}
where $\rho _{0s}=\gamma _{0s}^2(m_sn_{0s}+\Gamma _gP_{0s})$.\\
The continuity Eq. (\ref{eq15}), and Poisson's Eq. (\ref{eq16}) are linearized to obtain
\begin{eqnarray}
&\gamma _{0s}&\!\!\!\!\left(\frac{\partial }{\partial t}+u_{0s}\alpha \frac{\partial }{\partial z}+\frac{u_{0s}}{2r_+}+\gamma _{0s}^2\alpha \frac{du_{0s}}{dz}\right)\delta n_s+\left(\alpha \frac{\partial }{\partial z}+\frac{1}{2r_+}\right)(n_{0s}\gamma _{0s}u_{0s})\nonumber\\
&&+n_{0s}\gamma _{0s}^3\left[u_{0s}\frac{\partial }{\partial t}+\alpha \frac{\partial }{\partial z}+\frac{1}{2r_+}+\alpha \left(\frac{1}{n_{0s}}\frac{dn_{0s}}{dz}+3\gamma _{0s}^2u_{0s}\frac{du_{0s}}{dz}\right)\right]\delta u_s=0,\label{eq27}
\end{eqnarray}
and
\begin{eqnarray}
\frac{\partial \delta E_z}{\partial z}&=&4\pi e(n_{02}\gamma _{02}-n_{01}\gamma _{01})+4\pi e(\gamma _{02}\delta n_2-\gamma _{01}\delta n_1)+4\pi e(n_{02}u_{02}\gamma _{02}^3\delta u_2-n_{01}u_{01}\gamma _{01}^3\delta u_1).\label{eq28}
\end{eqnarray}
In similar fashion, the longitudinal part of the momentum conservation equation, Eq. (\ref{eq19}) is linearized to give
\begin{eqnarray}
\left\{\frac{\partial}{\partial t}+\alpha u_{0s}\frac{\partial}{\partial z}+\alpha \gamma ^2_{0s}(1+u^2_{0s})\frac{du_{0s}}{dz}\right\}\delta u_s-\frac{\alpha q_sn_{0s}}{\rho _{0s}\gamma _{0s}}\delta E_z+{\frac{1}{\gamma^2_{0s}n_{0s}}}\Bigg\{\frac{\gamma ^2_{0s}\gamma _gP_{0s}}{\rho_{0s}}\left(\alpha \frac{\partial}{\partial z}+u_{0s}\frac{\partial}{\partial t}\right)\nonumber\\
+\alpha \gamma ^2_{0s}\frac{\gamma _gP_{0s}}{\rho _{0s}}\left(\frac{1}{P_{0s}}\frac{dP_{0s}}{dz}-\frac{1}{n_{0s}}\frac{dn_{0s}}{dz}\right)+\left(1+\frac{\gamma ^2_{0s}\gamma _gP_{0s}}{\rho_{0s}}\right)\left(u_{0s}\gamma ^2_{0s}\alpha \frac{du_{0s}}{dz}+\frac{1}{2r_+}\right)\Bigg\}{\delta n_s}\nonumber\\
+\left(u_{0s}\alpha \frac{du_{0s}}{dz}+\frac{\alpha }{\rho _{0s}}\frac{dP_{0s}}{dz}+\frac{1}{\gamma ^2_{0s}2r_+}\right)= 0. \label{eq29}
\end{eqnarray}

\section{Local Approximation}\label{sec5}

In the paper \cite{one}, the unperturbed radial velocity, magnetic field, number density, and pressure near the event horizon for each species $s$ interms of  freefall velocity are given by
\begin{eqnarray}
u_{0s}(z)&=&[1-\alpha ^2(z)]^{\frac{1}{2}},\qquad B_0(z)=B_+v_{\rm ff}^4(z),\nonumber\\
P_{0s}(z)&=&P_{+s}v_{\rm ff}^{3\gamma _g}(z),\qquad n_{0s}(z)=\frac{1}{\zeta ^4}n_{+s}v_{\rm ff}^3(z)\label{eq30},
\end{eqnarray}
so that their derivatives become
\begin{eqnarray}
\frac{du_{0s}}{dz}&=&-\frac{\alpha }{2r_+}\frac{1}{v_{\rm ff}},\qquad \frac{dB_0}{dz}=-\frac{4\alpha }{2r_+}\frac{B_0}{v_{\rm ff}^2},\nonumber\\
\frac{dn_{0s}}{dz}&=&-\frac{3\alpha }{2r_+}\frac{n_{0s}}{v_{\rm ff}^2},\qquad \frac{dP_{0s}}{dz}=-\frac{3\alpha }{2r_+}\frac{\gamma _gP_{0s}}{v_{\rm ff}^2}\label{eq31},
\end{eqnarray}
where
\[
\zeta =\left[1+\frac{r_-}{r_+}\left(1-\frac{r_+}{r}\right)\right]^{\frac{1}{2}} \mbox{and} \quad v_{\rm ff}=[1-\alpha ^2(z)]^{1/2}.
\]

As mansion in paper \cite{one}, here we also applying the same restriction on a local scale for which the distance from the horizon does not vary significantly and use a local (or mean-field) approximation for the lapse function and hence for the equilibrium fields and fluid quantities. Since the plasma is situated relatively close to the horizon, i.e., $\alpha ^2\ll 1$, we choose a sufficiently small range in $z$ so that $\alpha $ does not vary much. In the local approximation for $\alpha $, $\alpha \simeq \alpha _o$ is valid within a particular layer. Hence, the unperturbed fields and fluid quantities, which are functions of $\alpha $, take on their corresponding \lq \lq mean-field\rq \rq values for a given $\alpha _0$.

Then, using the local approximation for $\alpha$, the derivatives of the equilibrium quantities can be evaluated at each layer for a given $\alpha_o$ of the form
\begin{eqnarray}
Du_{os}=\frac{du_{0s}}{dz}\Bigg|_{\alpha=\alpha_0}, \qquad Dn_{0s}=\frac{dn_{0s}}{dz}\Bigg|_{\alpha=\alpha_0},\nonumber\\
DP_{os}=\frac{dP_{0s}}{dz}\Bigg|_{\alpha=\alpha_0}, \qquad
DB_0 =\frac{dB_0}{dz}\Bigg|_{\alpha=\alpha_0}.\label{eq32}
\end{eqnarray}
The coefficients in Eqs. (\ref{eq27}), (\ref{eq28}), and (\ref{eq29}) then become constants within each layer with respect to $\alpha $ (and therefore $z$ as well). So it is possible to Fourier transform the equations with respect to $z$, using plane-wave-type solutions for the perturbations of the form $\sim e^{i(kz-\omega t)}$ for each $\alpha _0$ layer.

\section{Dispersion Relation}\label{sec6}

When Fourier transformed, the continuity Eq. (\ref{eq27}) with the help of Eq. (\ref{eq32}) becomes
\begin{equation}
\delta E={-}n_{0s}\gamma^2_{0s}\left[\frac{\alpha _0k-\omega u_{0s}-{\rm i}\alpha _0\left(Dn_{0s}/n_{0s}+3u_{0s}\gamma ^2_{0s}Du_{0s}\right)-{\rm i}/2r_+}{\alpha _0ku_{0s}-\omega -{\rm i}\left(u_{0s}/2r_++\alpha _0\gamma ^2_{0s}Du_{0s}\right)}\right]\delta u_s.\label{eq33}
\end{equation}
The longitudinal part of momentum equation, Eq. (\ref{eq29}) when  Fourier transformed, gives
\begin{eqnarray}
&&\left(\alpha_0ku_{0s}-\omega -{\rm i}\alpha_0\gamma ^2_{0s}(1+u^2_{0s})Du_{0s}\right)\delta u_s+\frac{{\rm i}\alpha_0q_sn_{0s}}{\rho _{0s}\gamma _{0s}}\delta E_z\nonumber\\
&&+{\frac{1}{\gamma^2_{0s}n_{0s}}}\left[\frac{\gamma ^2_{0s}\gamma _gP_{0s}}{\rho_{0s}}\left(\alpha_0k-\omega u_{0s}\right)-{\rm i}\alpha_0\gamma ^2_{0s}\frac{\gamma _gP_{0s}}{\rho _{0s}}\left(\frac{1}{P_{0s}}DP_{0s}-\frac{1}{n_{0s}}Dn_{0s}\right)-{\rm i}\Bigg(1+\frac{\gamma ^2_{0s}\gamma _gP_{0s}}{\rho_{0s}}\right)\nonumber\\
&&\left(\alpha _0\gamma ^2_{0s}u_{0s}Du_{0s}+\frac{1}{2r_+}\right)\Bigg]\delta n_s= 0.\label{eq34}
\end{eqnarray}
Finally, the Poisson's equation (\ref{eq28}) becomes
\begin{equation}
{\rm i}k\delta E_z=4\pi e(\gamma _{02}\delta n_2-\gamma _{01}\delta n_1)+4\pi e(n_{02}u_{02}\gamma^3_{02}\delta u_2-n_{01}u_{01}\gamma^3_{01}\delta u_1). \label{eq35}
\end{equation}
Using Eqs. (\ref{eq33}), (\ref{eq34}), and (\ref{eq35}), the dispersion relation for the longitudinal wave modes may be put in the form
\begin{eqnarray}
&&1=\frac{\alpha _0}{k}\sum_s\frac{\omega^2_{ps}}{\gamma^2_{0s}}\left[\alpha _0k-\frac{{\rm i}}{2r_+}-{\rm i}\alpha_0\gamma^2_{0s}\left(\frac{1}{n_{0s}}Dn_{0s}+2\gamma^2_{0s}u_{0s}Du_{0s}\right)\right]\nonumber\\
&&\times\Bigg(\left[\alpha_0ku_{0s}-\omega -{\rm i}\alpha_0\gamma ^2_{0s}(1+u^2_{0s})Du_{0s}\right]\left(\alpha_0ku_{0s}-\omega -{\rm i}\alpha_0\gamma ^2_{0s}Du_{0s}-\frac{{\rm i}u_{0s}}{2r_+}\right)\nonumber\\
&&-\frac{v^2_{Ts}}{2}\left[\alpha _0k-\omega u_{0s}-\frac{{\rm i}}{2r_+}-{\rm i}\alpha_0\left(\frac{1}{n_{0s}}Dn_{0s}+3\gamma^2_{0s}u_{0s}Du_{0s}\right)\right]\nonumber\\
&&\times\left[\alpha_0k-\omega u_{0s}-{\rm i}\alpha_0\left(\frac{1}{P_{0s}}DP_{0s}-\frac{1}{P_{0s}}DP_{0s}\right)-\frac{2{\rm i}}{v^2_{Ts}}\left(1+\frac{v^2_{Ts}}{2}\right)\left(\frac{1}{2r_+}+\alpha_0\gamma^2_{0s}u_{0s}Du_{0s}\right)\right]\Bigg)^{-1},\quad \quad\label{eq36}
\end{eqnarray}
where $v^2_{Ts}=2\gamma _g\gamma^2_{0s}P_{0s}/\rho_{0s}$ is the thermal velocity, $\omega _{cs}=e\gamma _{0s}n_{0s}B_0/\rho _{0s}$ the cyclotron frequency, and $\omega^2 _{ps}=4\pi e^2\gamma^2_{0s}n^2_{0s}/\rho _{0s}$ the plasma frequency of the fluid\rq s of species $s$.\\

Note that, the plasma frequency is also frame independent, that is, independent of $\gamma _{0s}$  since the $\gamma^2 _{0s}$   factor in the numerator cancel out the $\gamma ^2_{0s}$  factor involved the energy density, $\rho _{0s}$  , in the denominator. The same is also true for the thermal velocity, for which the $\gamma ^2_{0s}$   factor cancels. The above dispersion relation is valid for either electron-positron or electron-ion plasma in that it makes no assumption as to the mass, number density, or temperature of each species. In this respect it is completely general. If one considers the equivalent case to that of SK \cite{sixteen} in Eq. (\ref{eq36}) for an electron-positron plasma, in which the two fluids have the same velocity $u_0$ , the same equilibrium density $n_0$, and at the same temperature $T_0$, one obtains
\begin{eqnarray}
&&1=\frac{2\omega^2_p\alpha _0}{\gamma^2_0k}\left[\alpha _0k-\frac{{\rm i}}{2r_+}-{\rm i}\alpha_0\gamma^2_{0}\left(\frac{1}{n_{0}}Dn_{0}+2\gamma^2_{0}u_{0}Du_{0}\right)\right]\nonumber\\
&&\times\Bigg\{\left[\alpha_0ku_{0}-\omega -{\rm i}\alpha_0\gamma ^2_{0}(1+u^2_{0})Du_{0}\right]\left(\alpha_0ku_{0}-\omega -{\rm i}\alpha_0\gamma ^2_{0}Du_{0}-\frac{{\rm i}u_{0}}{2r_+}\right)\nonumber\\
&&-\frac{v^2_{T}}{2}\left[\alpha _0k-\omega u_{0}-\frac{{\rm i}}{2r_+}-{\rm i}\alpha_0\left(\frac{1}{n_{0}}Dn_{0}+3\gamma^2_{0}u_{0}Du_{0}\right)\right]\nonumber\\
&&\times\left[\alpha_0k-\omega u_{0}-{\rm i}\alpha_0\left(\frac{1}{P_{0}}DP_{0}-\frac{1}{P_{0s}}DP_{0s}\right)-\frac{2{\rm i}}{v^2_{T}}\left(1+\frac{v^2_{T}}{2}\right)\left(\frac{1}{2r_+}+\alpha_0\gamma^2_{0}u_{0}Du_{0}\right)\right]\Bigg\}^{-1}.\label{eq37}
\end{eqnarray}
In the limit of zero gravity, i.e., $\alpha \rightarrow 1$ and with the velocities  $u_0\rightarrow 0$, Lorentz factor $\gamma _0\rightarrow 1$, the derivative of unperturbed quantities  $Du_0=Dn_0=DP_0=DB_0=0$, it reduced to
$\omega ^2=\left(2\omega ^2_p+k^2\gamma _gP_0/\rho_0\right)$, which recovered the SK \cite{sixteen} result. The only difference being that  $\gamma _g $ was set to unity in the SK work.

\section{Numerical Solution of Longitudinal Modes}\label{sec7}

The dispersion relation given in Eq. (\ref{eq36}) is complicated enough, even in the simplest cases for the electron-positron plasma where both species are assumed to have the same equilibrium parameters, and an analytical solution is cumbersome and unprofitable. We solve numerically the dispersion relation in order to determine all the physically meaningful modes for the transverse waves. We put the equations in the form of a matrix equation as follows:
\begin{equation}
(A-kI)X=0.\label{eq38}
\end{equation}
The eigenvalue is chosen to be the wave number $k$, the eigenvector $X$ is given by the relevant set of perturbations, and $I$ is the identity matrix.

We need to write the perturbation equations in an appropriate form. We introduce the following set of dimensionless variables:
\begin{eqnarray}
&&\tilde \omega =\frac{\omega }{\alpha _0\omega _\ast },\quad \tilde k=\frac{kc}{\omega _\ast },\quad k_+=\frac{1}{2r_+\omega _\ast },\nonumber\\
&&\delta \tilde u_s=\frac{\delta u_s}{u_{0s}},\quad \tilde v_s=\frac{\delta v_s}{u_{0s}},\quad \delta \tilde n_s=\frac{\delta n_s}{n_{0s}},\nonumber\\
&&\delta \tilde B=\frac{\delta B}{B_0},\quad \tilde E=\frac{\delta E}{B_0},\quad \delta \tilde E_z=\frac{\delta E_z}{B_0}.\label{eq39}
\end{eqnarray}
The derivatives of the equilibrium quantities given in the equation \ref{eq45} are also expressed in terms of $k_+$ so that
\begin{equation}
\frac{dv_{\rm ff}}{dz}\Bigg|_{\alpha =\alpha _0}={-}\frac{\alpha_0k_+}{v_{\rm ff}},\label{eq40}
\end{equation}
and
\begin{eqnarray}
D\tilde u_{0s}&=&-\frac{\alpha_0k_+}{v_{\rm ff}},\qquad \qquad D\tilde B_0=-\frac{4\alpha_0k_+}{v_{\rm ff}^2}B_0,\nonumber\\
D\tilde n_{0s}&=&-\frac{3\alpha_0k_+}{v_{\rm ff}^2}n_{0s},\qquad D\tilde P_{0s}=-\frac{3\alpha_0k_+\gamma _g}{v_{\rm ff}^2}P_{0s}\label{eq41}.
\end{eqnarray}
For an electron-positron plasma, $\omega _{p1}=\omega _{p2}$ and $\omega _{c1}=\omega _{c2}$, and for electron-ion plasma $\omega _p=\sqrt{\omega _{p1}\omega _{p2}}$ , $\omega _c=\sqrt{\omega _{c1}\omega _{c2}}$,
and $\omega _{\ast s}^2=(2\omega _{ps}^2+\omega _{cs}^2)$. Here, $\omega _\ast $ is defined as
\begin{equation}
\omega _\ast =\omega_{P}=\sqrt(\omega_{p1}\omega_{p2}).\label{eq42}
\end{equation}
The dimensionless eigenvector for the Longitudinal set of equations is
\begin{equation}
\tilde X_{\rm Longitudinal}=\left[\begin{array}{c}\delta \tilde u_1\\\delta \tilde u_2\\\delta \tilde n_1\\\delta \tilde n_2\\\delta \tilde E_z\end{array}\right]\label{eq43}.
\end{equation}
Using Eqs. (\ref{eq39}) and (\ref{eq41}), the Eqs. (\ref{eq33}), (\ref{eq34}), and (\ref{eq35}) can be written in the following dimensionless form
\begin{eqnarray}
\tilde k\delta\tilde u_s=\frac{1}{u^2_{0s}-v^2_{Ts}/2}\left\{u_{0s}\tilde\omega \left(1-\frac{v^2_{Ts}}{2}\right)-{\rm i}\frac{k_+}{\alpha _0}\left[\left(u^2_{0s}-v^2_{Ts}/2\right)\left(1+\frac{1}{u^2_{0s}}\right)-\frac{v^2_{Ts}}{u^2_{0s}}\right]\right\}\delta \tilde u_s\nonumber\\
-\frac{1}{\gamma^2_{0s}}\frac{1}{u^2_{0s}-v^2_{Ts}/2}\frac{v^2_{Ts}\tilde\omega}{2u_{0s}\gamma^2_{0s}}\delta\tilde n_s-\frac{{\rm i}(q_s/e)\omega_{cs}}{\gamma^2_{0s}\omega_p({u^2_{0s}-v^2_{Ts}/2})}\delta\tilde E_z, \label{eq44}
\end{eqnarray}
\begin{eqnarray}
\tilde k\delta\tilde n_s=\frac{1}{u^2_{0s}-v^2_{Ts}/2}\left\{u_{0s}\tilde\omega \left(1-\frac{v^2_{Ts}}{2}\right)-{\rm i}\frac{k_+\alpha _0}{u^2_{0s}}\left(u^2_{0s}-v^2_{Ts}/2\right)\right\}\delta \tilde n_s\nonumber\\
-\frac{u^2_{0s}\gamma^2_{0s}}{u^2_{0s}-v^2_{Ts}/2}\left(\frac{\tilde\omega}{u_{0s}\gamma^2_{0s}}+{\rm i}2\gamma_{0s}k_+\right)\delta\tilde u_s+\frac{{\rm i}(q_s/e)\omega_{cs}}{\omega_p({u^2_{0s}-v^2_{Ts}/2})}\delta\tilde E_z, \label{eq45}
\end{eqnarray}
\begin{eqnarray}
\tilde k\delta \tilde E_z&=&{\rm i}u^2_{01}\gamma^2_{01}\frac{\omega _{p1}^2}{\omega _{c1}\omega _p }\delta \tilde u_1-{\rm i}u^2_{02}\gamma^2_{02}\frac{\omega _{p2}^2}{\omega _{c2}\omega _p }\delta \tilde u_2+{\rm i}\frac{\omega _{p1}^2}{\omega _{c1}\omega _p }\delta \tilde n_1-{\rm i}\frac{\omega _{p2}^2}{\omega _{c2}\omega _p }\delta \tilde n_2\label{eq46}.
\end{eqnarray}
These are the equations in the required form to be used as input to Eq. (\ref{eq36}).
Using Eqs. (\ref{eq44})-(\ref{eq46}), the Eq. (\ref{eq38}) then can be written as
\begin{equation}
(\tilde A-\tilde kI)\tilde X=0.\label{eq47}
\end{equation}
The numerical solution is carried out using  MATLAB as performed in paper \cite{one} with $\zeta ^2=1.05$ and $q^2/M^2=0.2$. The limiting horizon values for the electron-positron plasma and electron-ion plasma are taken respectively as
\begin{equation*}
n_{+s}=10^{18}{\rm cm}^{-3},\;T_{+s}=10^{10}{\rm K},\;B_+=3\times 10^6{\rm G},\;\mbox{and}\;\gamma _g=\frac{4}{3},
\end{equation*}
\begin{equation*}
n_{+1}=10^{18}{\rm cm}^{-3},\;T_{+1}=10^{10}{\rm K},\;n_{+2}=10^{15}{\rm cm}^{-3},\;T_{+2}=10^{12}{\rm K},
\end{equation*}
with  $\gamma _g=\frac{4}{3},\;M=5M_\odot ,\;q^2=0.2M^2$.
\begin{figure}[h]
\begin{center}
 \includegraphics[scale=0.4]{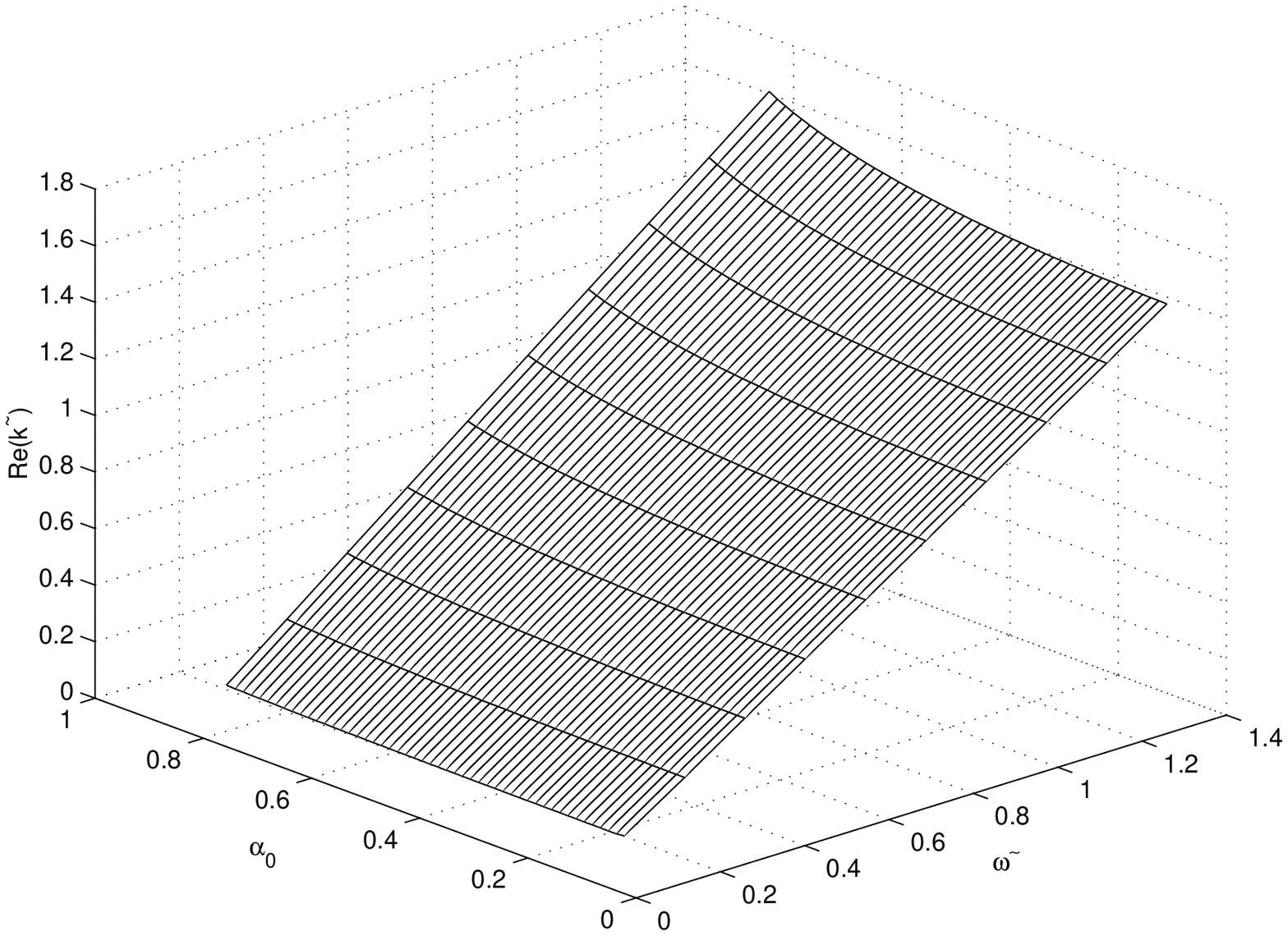}
 \includegraphics[scale=0.4]{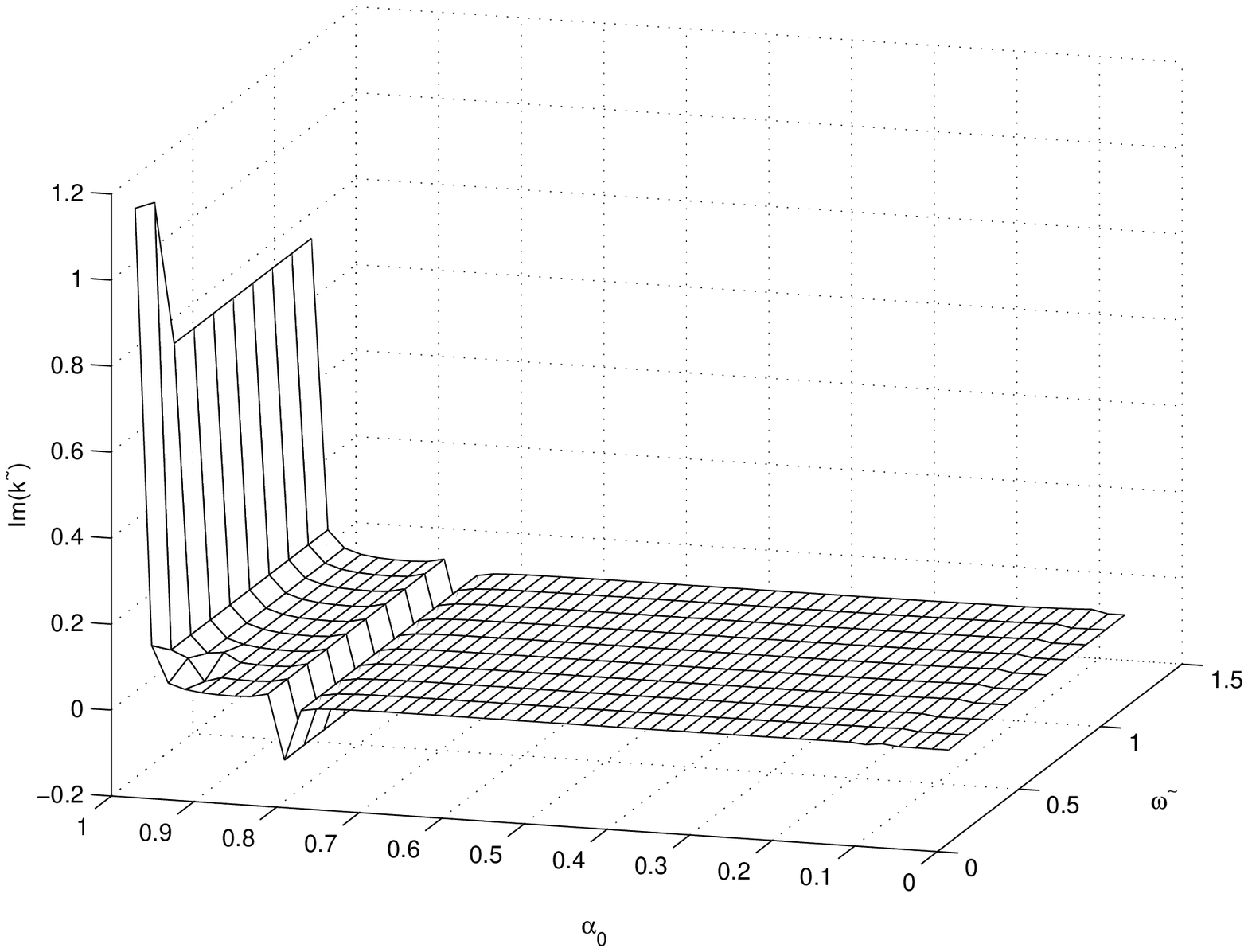}
\end{center}
\caption{\it Longitudinal low frequency mode showing both damping and growth for electron-positron plasma.}\label{fig1}
\begin{center}
 \includegraphics[scale=0.4]{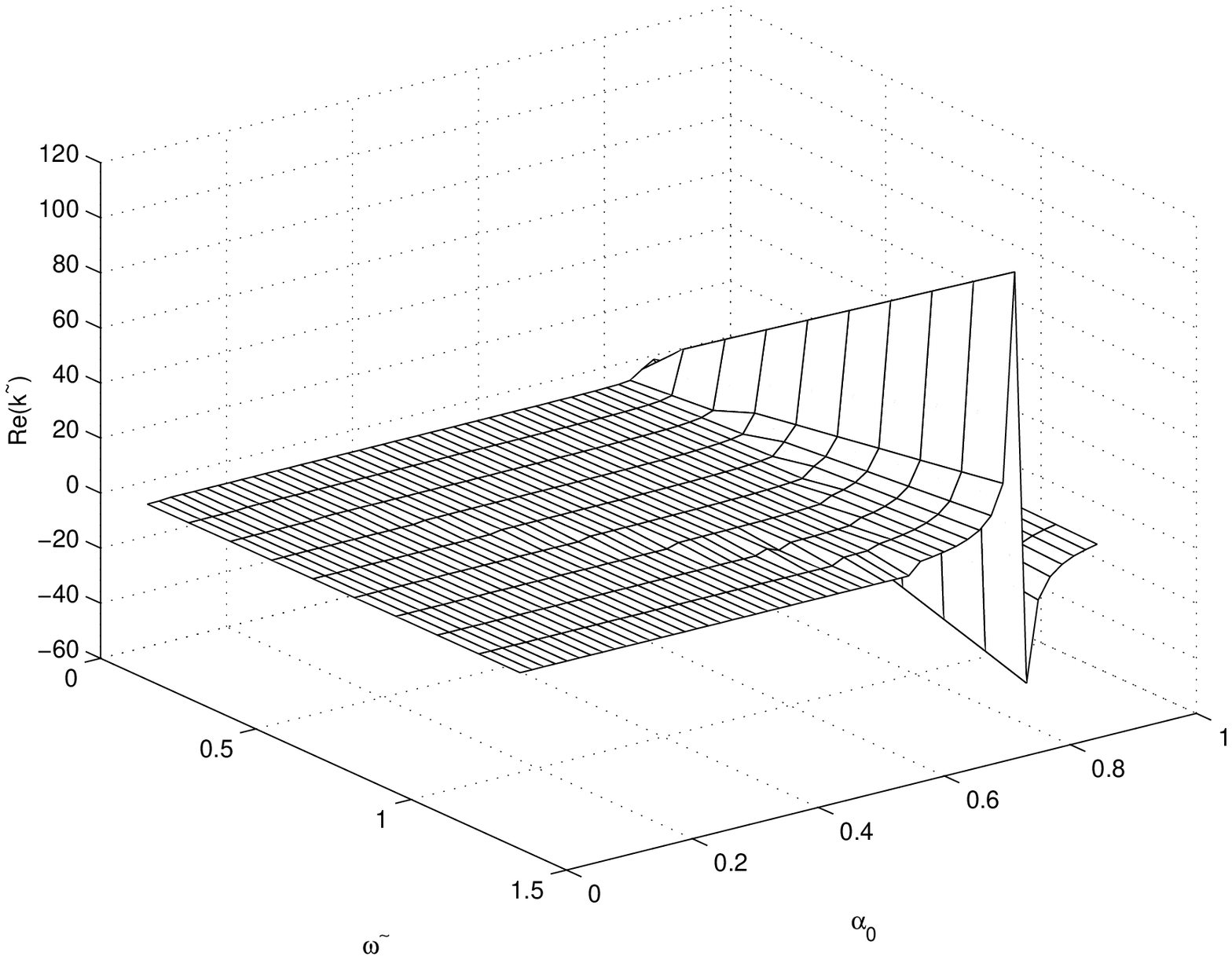}
 \includegraphics[scale=0.4]{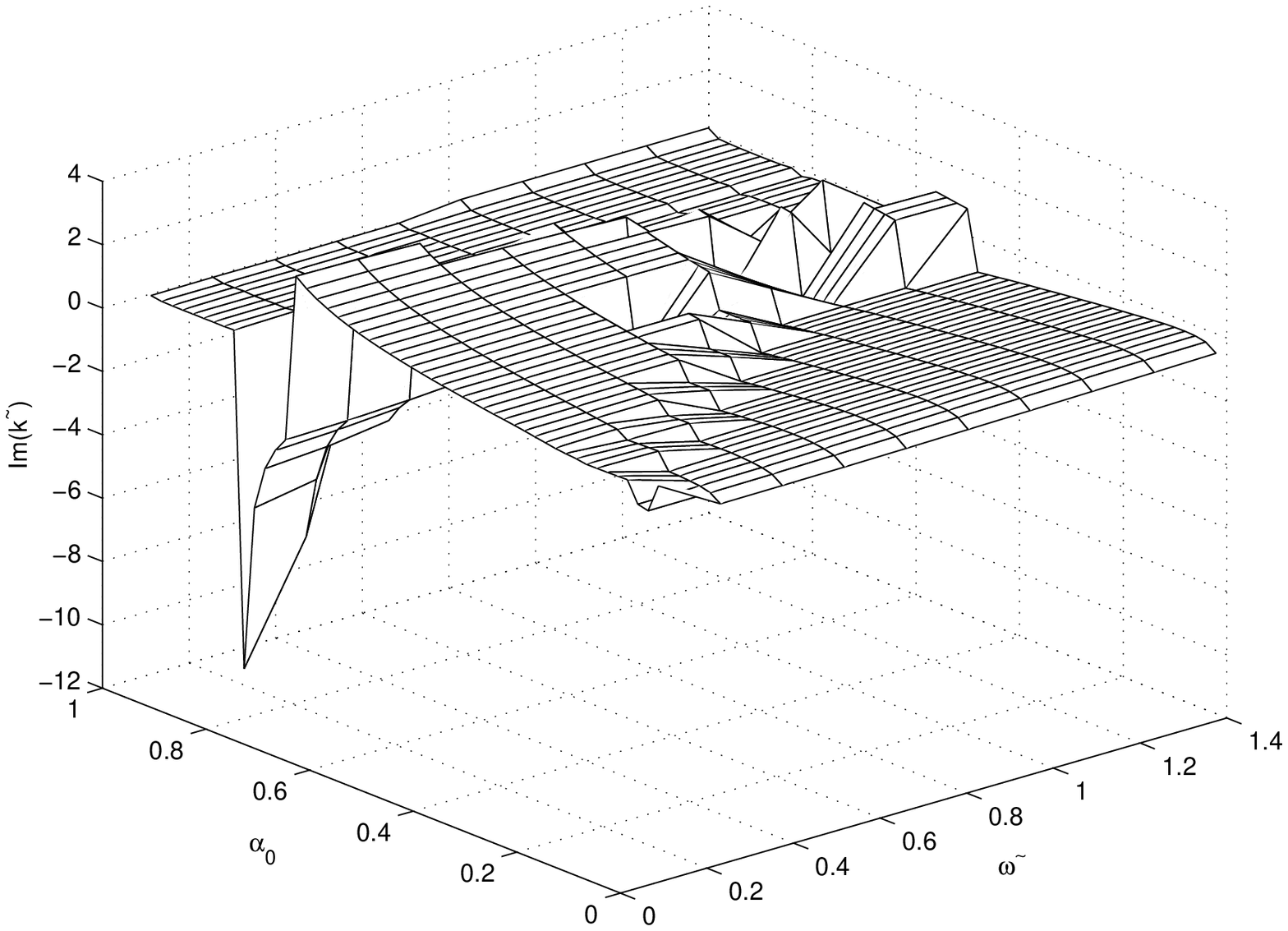}
\end{center}
\caption{\it This mode shows some interplay between frequency and distance from the horizon $\alpha_0$ for electron-positron plasma.}\label{fig2}
\end{figure}

\section{Results}\label{sec8}

\subsection{Electron-positron Plasma}\label{subsubsec8.1}
\centerline{\large{\bf 1. Longitudinal Low Frequency Mode}}
The longitudinal modes for electron-positron plasma are split into high and low frequency modes with domain as $\tilde\omega <\sqrt 2$ and $\tilde\omega >\sqrt 2$. For low frequency our study show five modes to exist, while for Schwarzschild case four low frequency modes are found \cite{fifteen}. These are in contrast with the special relativistic case investigated by SK \cite{sixteen} where only one high frequency mode was found to exist for electron-positron plasma.
\begin{figure}[h]
\begin{center}
 \includegraphics[scale=.4]{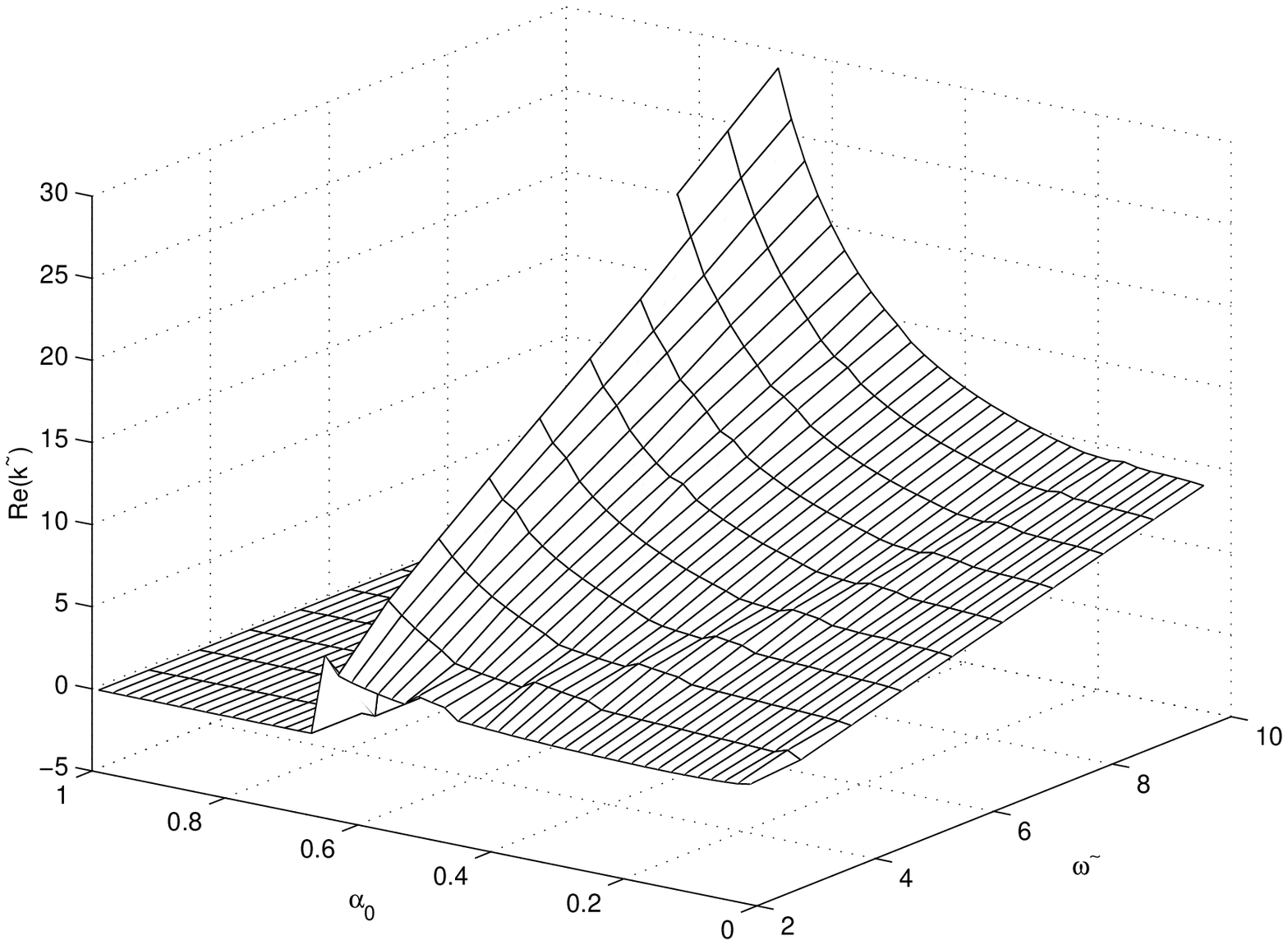}
 \includegraphics[scale=.4]{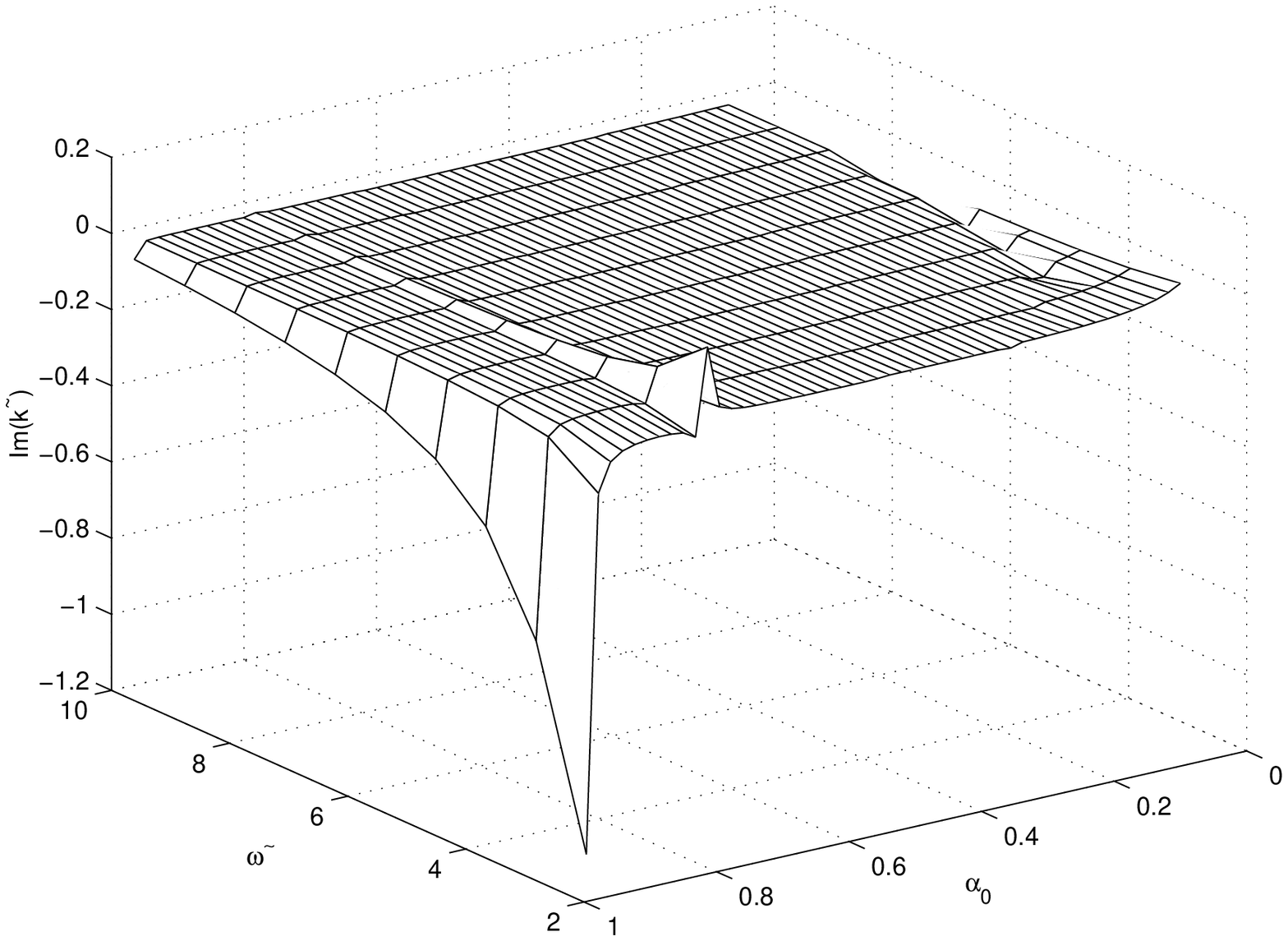}
\end{center}
\caption{\it Longitudinal high frequency growing and damped mode for electron-positron plasma.}\label{fig3}
\begin{center}
 \includegraphics[scale=.4]{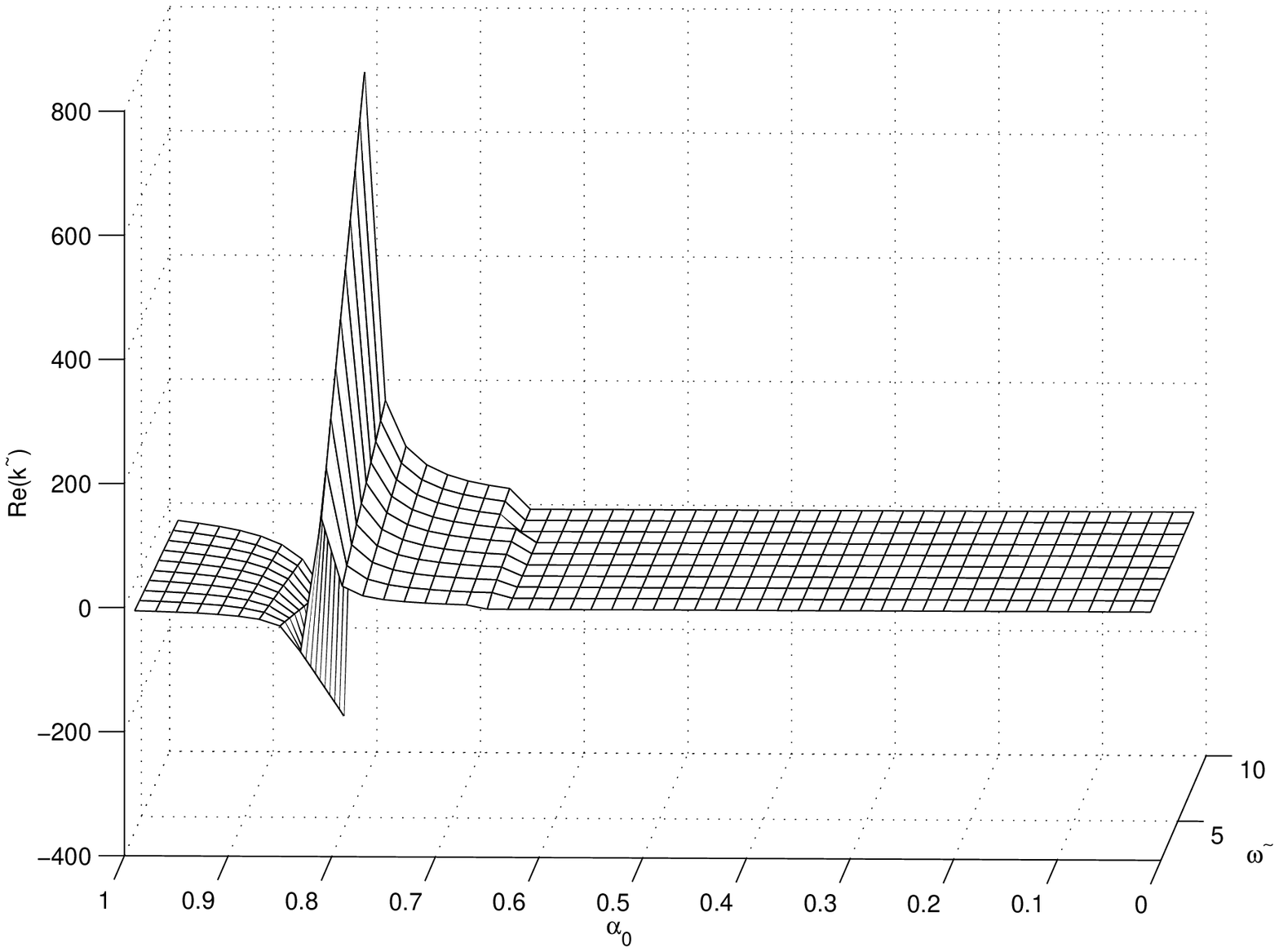}
 \includegraphics[scale=.4]{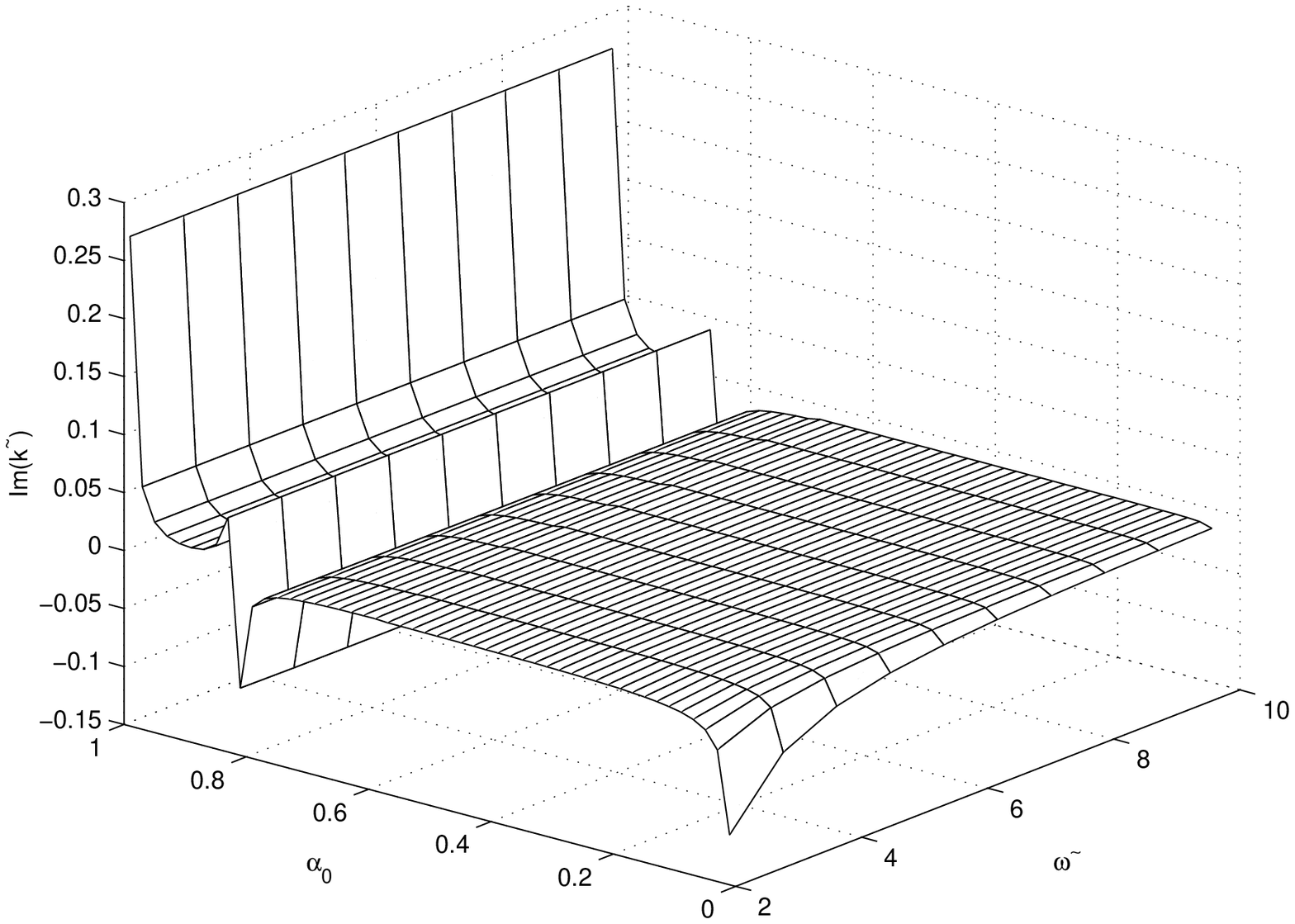}
\end{center}
\caption{\it Longitudinal high frequency damping and growing mode for electron-positron plasma.}\label{fig4}
\end{figure}
The first low frequency mode shown in Fig. 1 is physical and shows growth for $\alpha _0<\alpha _t$ but is damped for $\alpha _0>\alpha _t$, where $\alpha _t$ denotes the transonic radius occurs at about $\alpha _t\sim 0.82$ for the case considered here. This means that energy is drained from the wave rather than being fed into it by the gravitational field. The second low frequency mode is similar to the mode shown in Fig. 1. The third low frequency mode shown in Fig. 2 is also physical for $\alpha _0<\alpha _t$. This mode is odd in that there appears to be some interplay between frequency and distance from the horizon $\alpha_0$, which split this mode into two distinct regions. One is purely growth region for which $Im(k)>>Re(k)$ and the other which is also a growth region but where $Im(k)<< Re(k)$. The fourth low frequency mode is opposite in sense to the previous mode and is not shown here. The fifth and final low frequency mode has a little difference from the mode shown in Fig. 2 and it is also true for this mode that there appears some interplay between frequency and distance from the horizon $\alpha_0$. It is evident that the growth and damping rates are independent of the frequency, but depended only on the radial distance from the black hole horizon through $\alpha _0$.
\begin{figure}[h]
\begin{center}
 \includegraphics[scale=.4]{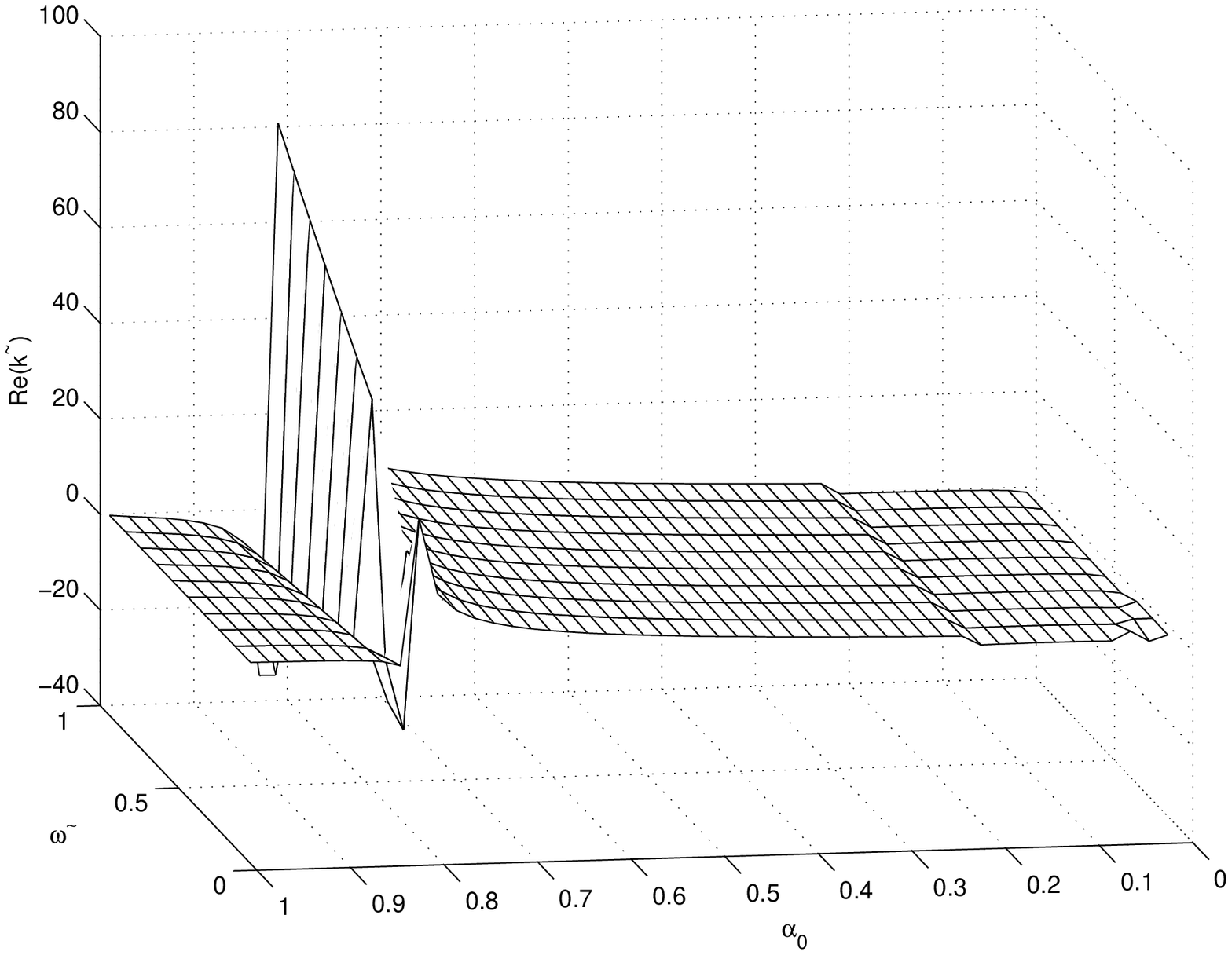}
 \includegraphics[scale=.4]{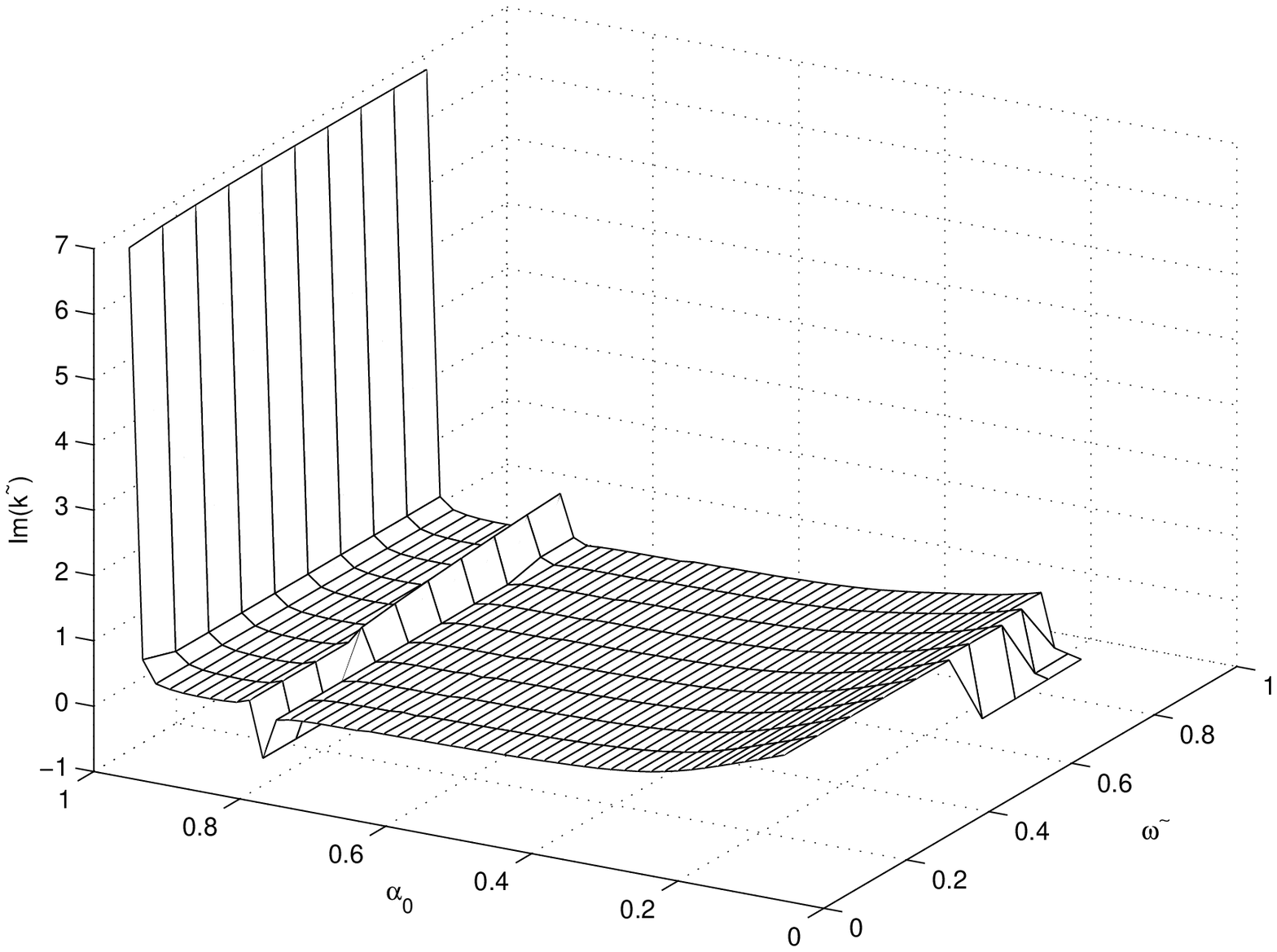}
\end{center}
\caption{\it Longitudinal low frequency damped mode for electron-ion plasma.}\label{fig5}
\begin{center}
 \includegraphics[scale=.4]{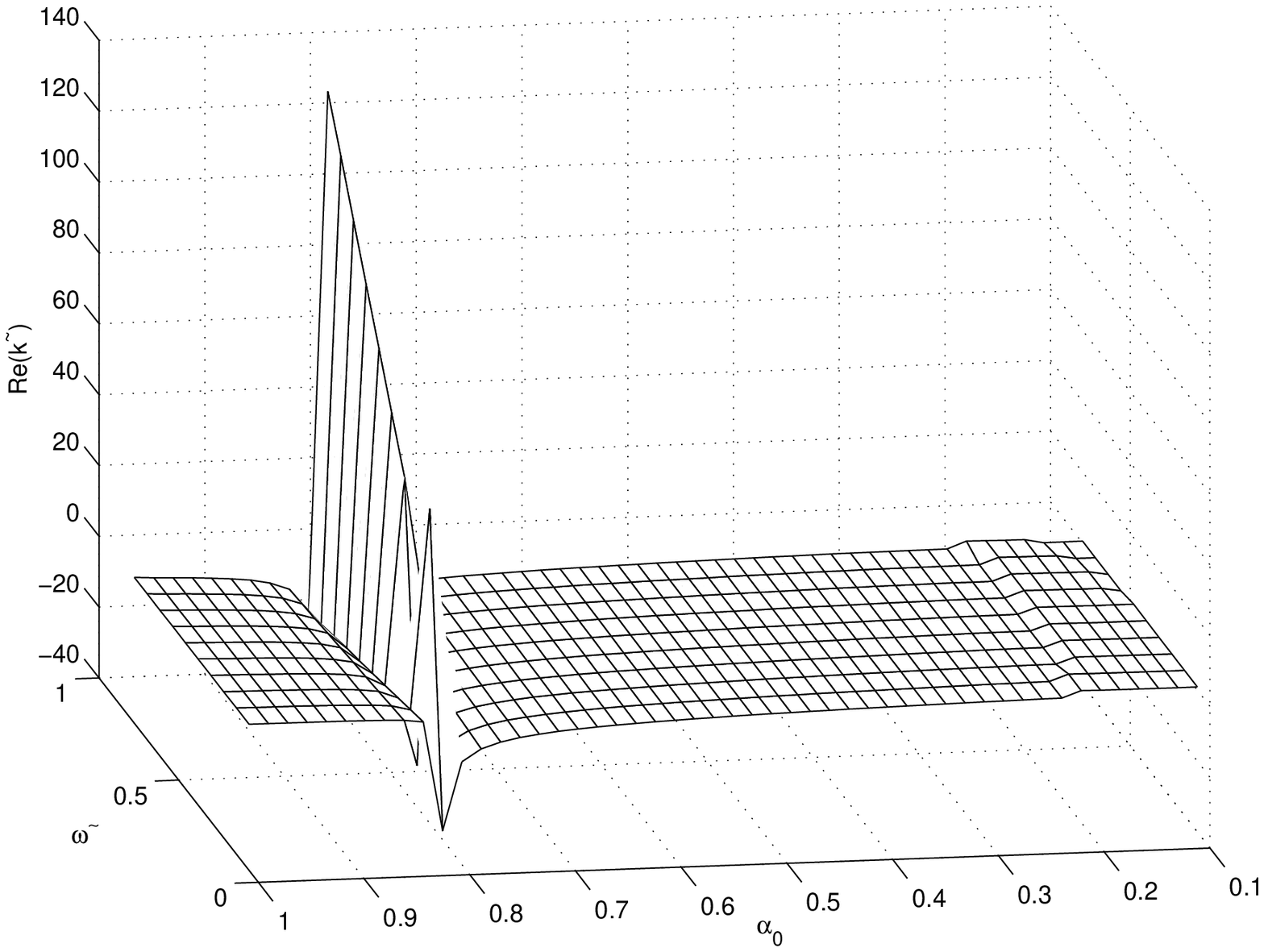}
 \includegraphics[scale=.4]{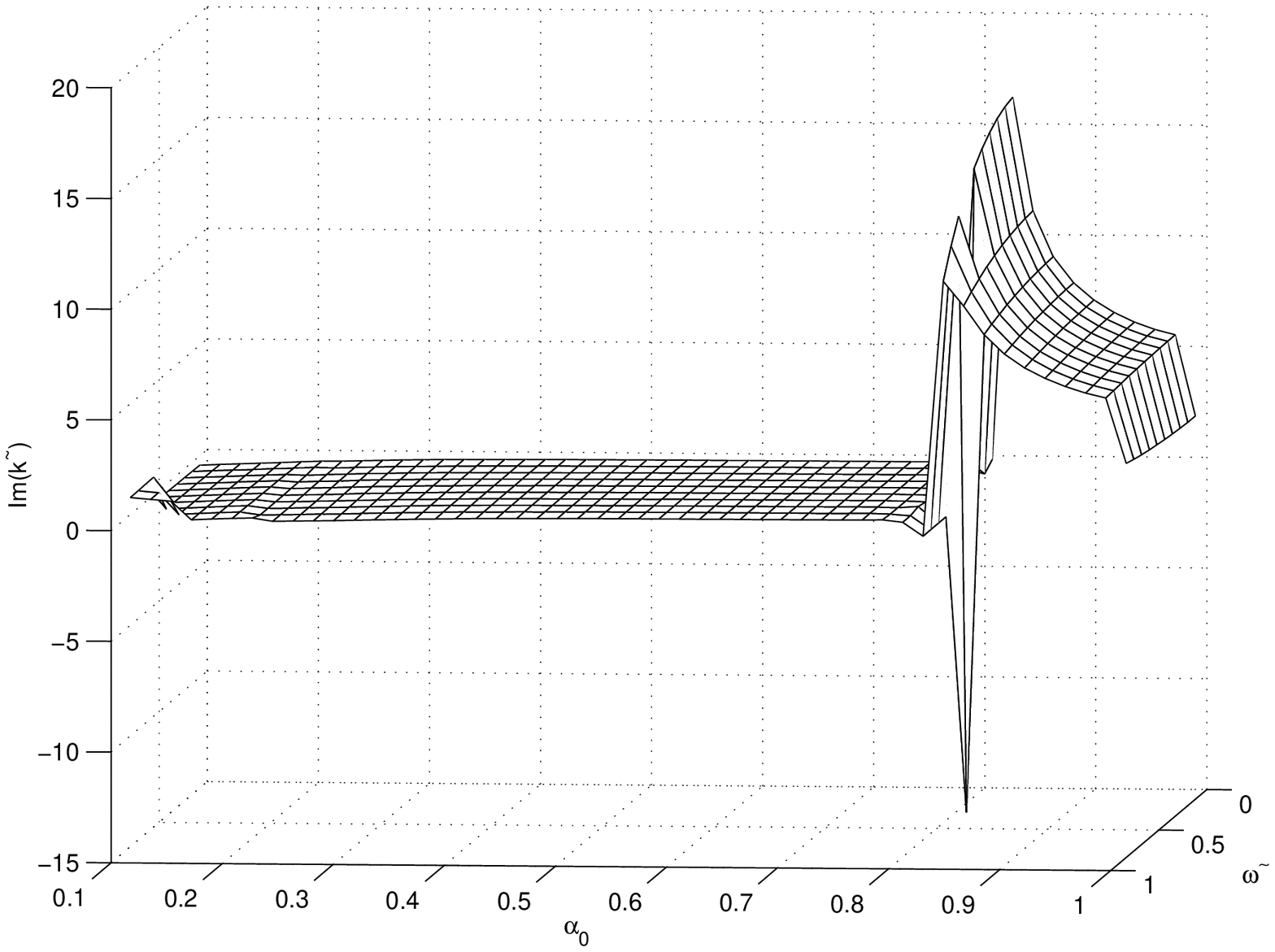}
\end{center}
\caption{\it Longitudinal low frequency growth mode for electron-ion plasma.}\label{fig6}
\end{figure}

\centerline{\large{\bf 2. Longitudinal High Frequency Mode}}
In the high frequency domain there exist five modes. The first two modes are equivalent to the modes shown in Fig. 1 and show growth for $\alpha _0<\alpha _t$ and damped for $\alpha _0>\alpha _t$. Again, energy is drained from the wave rather than being fed into it by the gravitational field. The third mode shown in Fig. 3 is a new mode which shows damped corresponding to $\alpha _0<\alpha _t$ and growth for $\alpha _0>\alpha _t$. The other two modes (one is depicted in Fig. 4) also show growth for $\alpha _0<\alpha _t$ and damped above $\alpha _t $.\\

\subsection{Electron-ion Plasma}\label{subsubsec8.2}
\centerline{\large{\bf 1. Longitudinal Low Frequency Mode}}
For electron-ion plasma the longitudinal modes are split into high and low frequency domain as $\tilde\omega <1$ for low frequency, and $\tilde\omega>1$ for high frequency. As for electron-positron case, here we have found five low frequency modes to exist. Since the temperature of electron and ion near the horizon are different so their transonic radius is same for-electron positron case occurs at $\alpha _{t1}\sim 0.82$ . The transonic radius for ion occurs at $\alpha _{t2}>\sim 0.99$ and can not properly be shown for low frequency modes. The first two modes are similar to the modes shown in Fig. 4 for electron-positron plasma and both show growth for $\alpha _0<\alpha _{t1}$ and damped for $\alpha _0>\alpha _{t1}$. This is interesting again that energy is being fed into the wave between the transonic radii but is drained from the wave closed to the horizon. The third mode illustrated in Fig. 5 is damped for $0<\alpha_0<1$. The remaining two modes are equivalent to one another and show growth for $\alpha _0<\alpha _{t1}$ and for $\alpha_{t1}<\alpha _0<\alpha _{t2}$ and only one of these modes is shown in Fig. 6. It is clear that the growth and damping rates are frequency independent as for electron-positron plasma, but depended only on the distance from the black hole horizon through $\alpha _0$.
\begin{figure}[h]
\begin{center}
 \includegraphics[scale=.4]{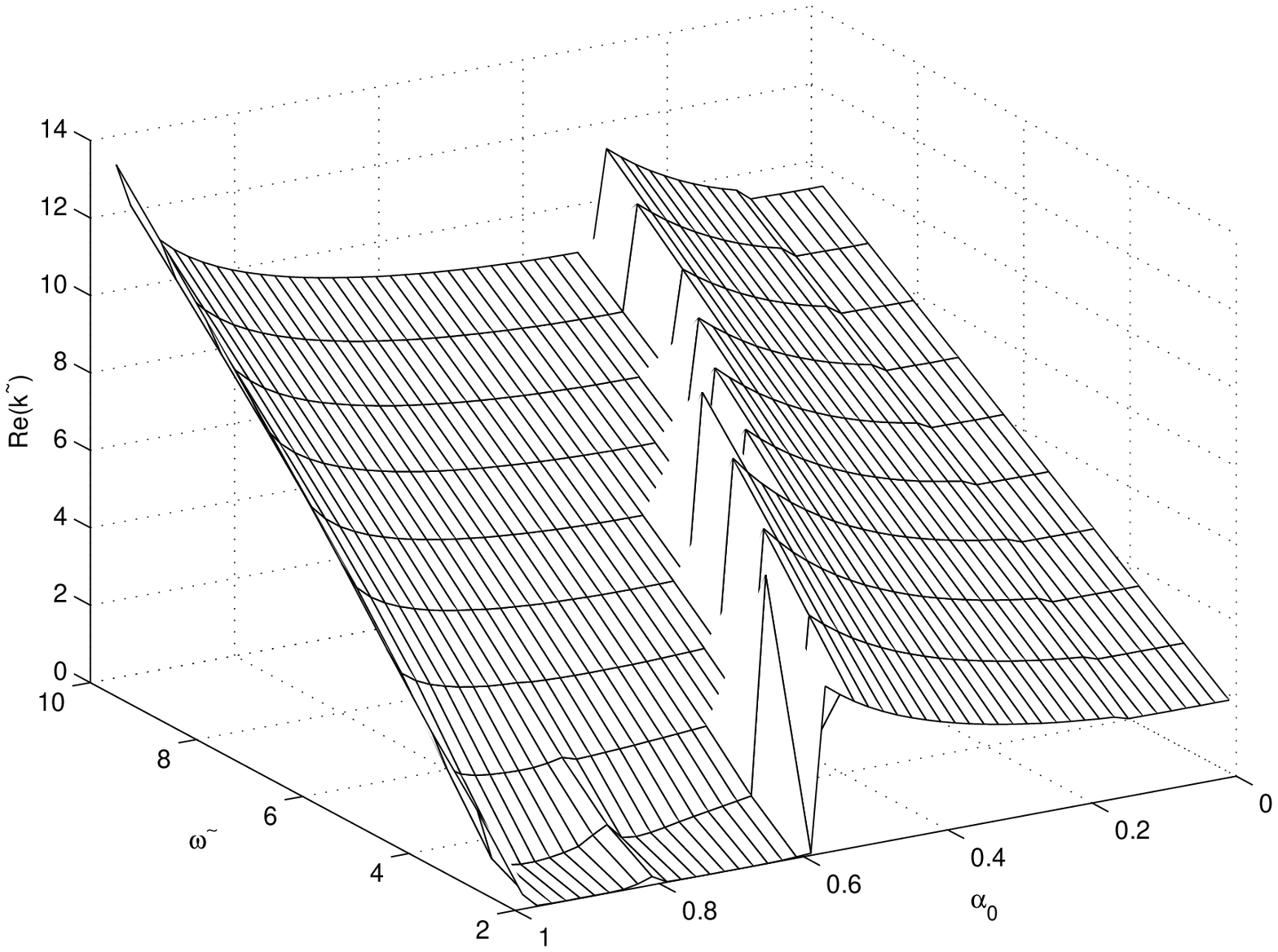}
 \includegraphics[scale=.4]{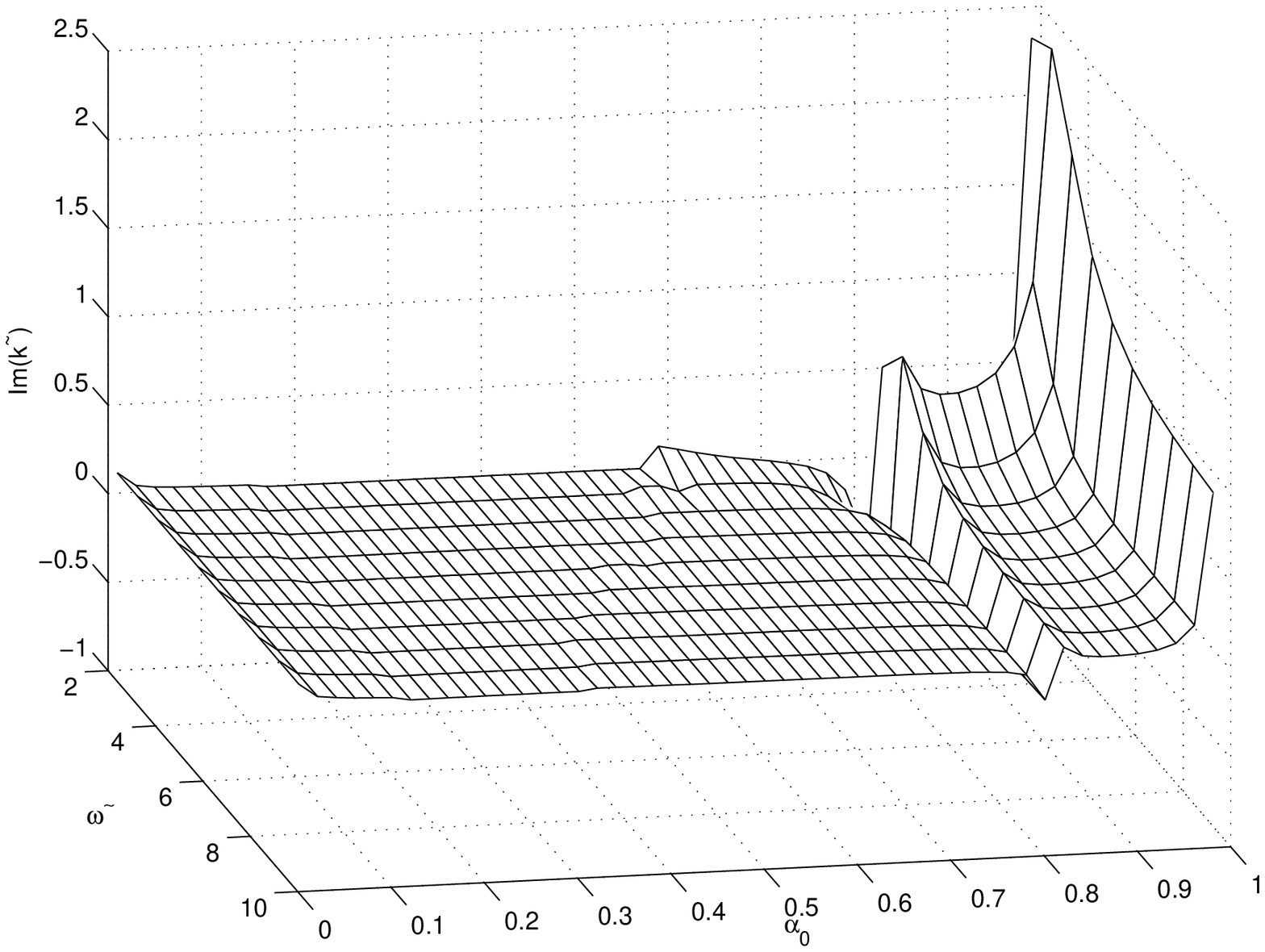}
\end{center}
\caption{\it Longitudinal high frequency mode both damping and growth for electron-ion plasma.}\label{fig7}
\end{figure}

\centerline{\large{\bf 2. Longitudinal High Frequency Mode}}
Four high frequency modes are found to exist as shown in Figs. 7 and 8.
The first mode not shown here is similar to the first high frequency mode shown in Fig. 1 for electron-positron plasma. The second mode shown in Fig. 7 is also growing for $\alpha _0<\alpha _{t1}$ and damped for $\alpha _0>\alpha _{t1}$. There appear some interplay between frequency and lapse function $\alpha_0$. Thus near the transonic radius, it appears that energy is no longer fed into wave mode by the gravitational field but begins to be drained from the waves. Fig. 8 shows two modes, one of which shown in bottom left is growing for $\alpha _0<\alpha _{t1}$ and damped for $\alpha _0>\alpha _{t1}$, while the bottom right one is damped for $\alpha _0<\alpha _{t2}$ and growth for $\alpha _0>\alpha _{t2}$. The transonic radius for ions is clearly shown here. The instability (damped and growing) of longitudinal wave modes has occurred at each transonic radius and is interesting.
\begin{figure}[h]
\begin{center}
 \includegraphics[scale=.4]{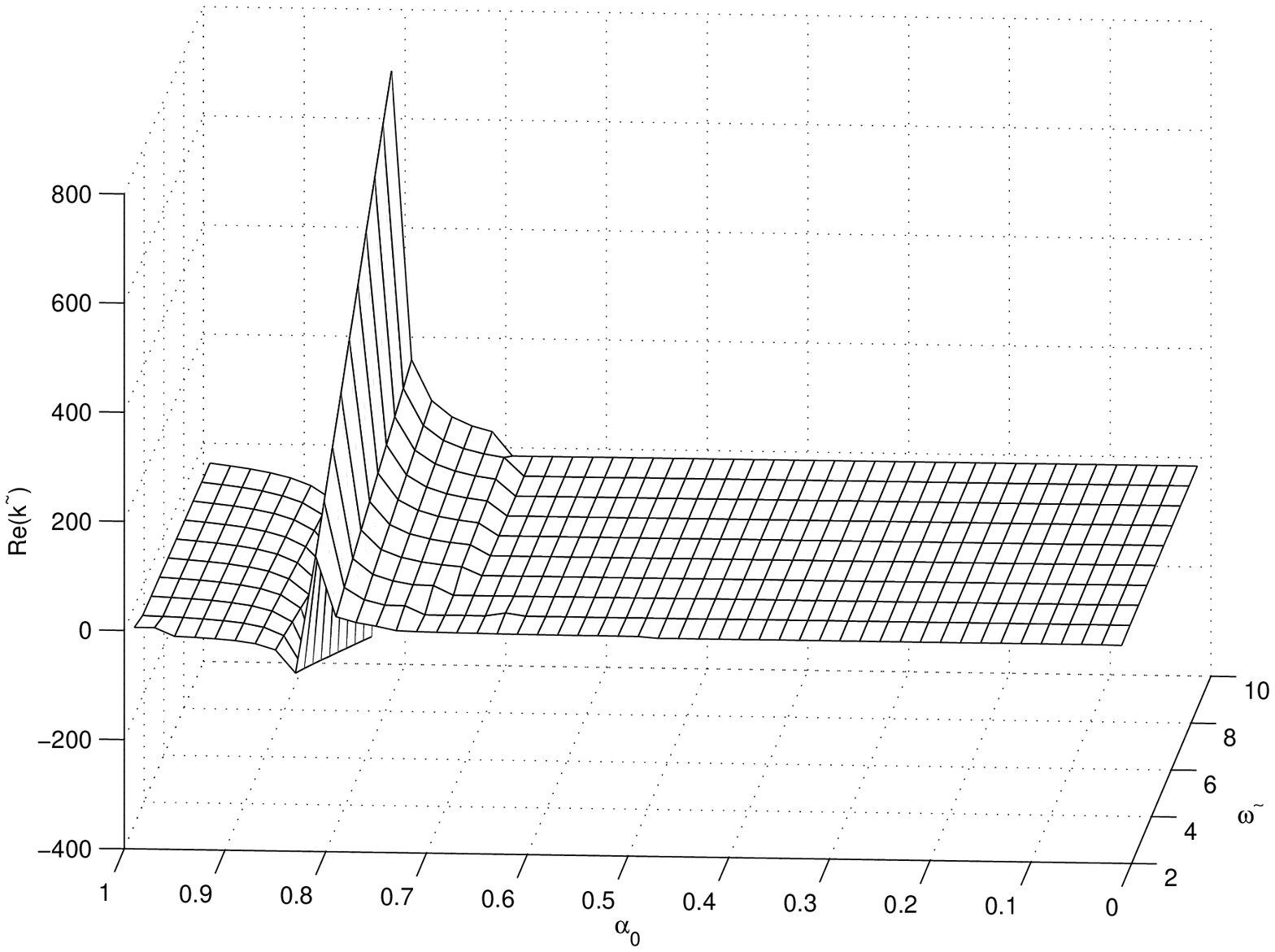}\\
 \includegraphics[scale=.4]{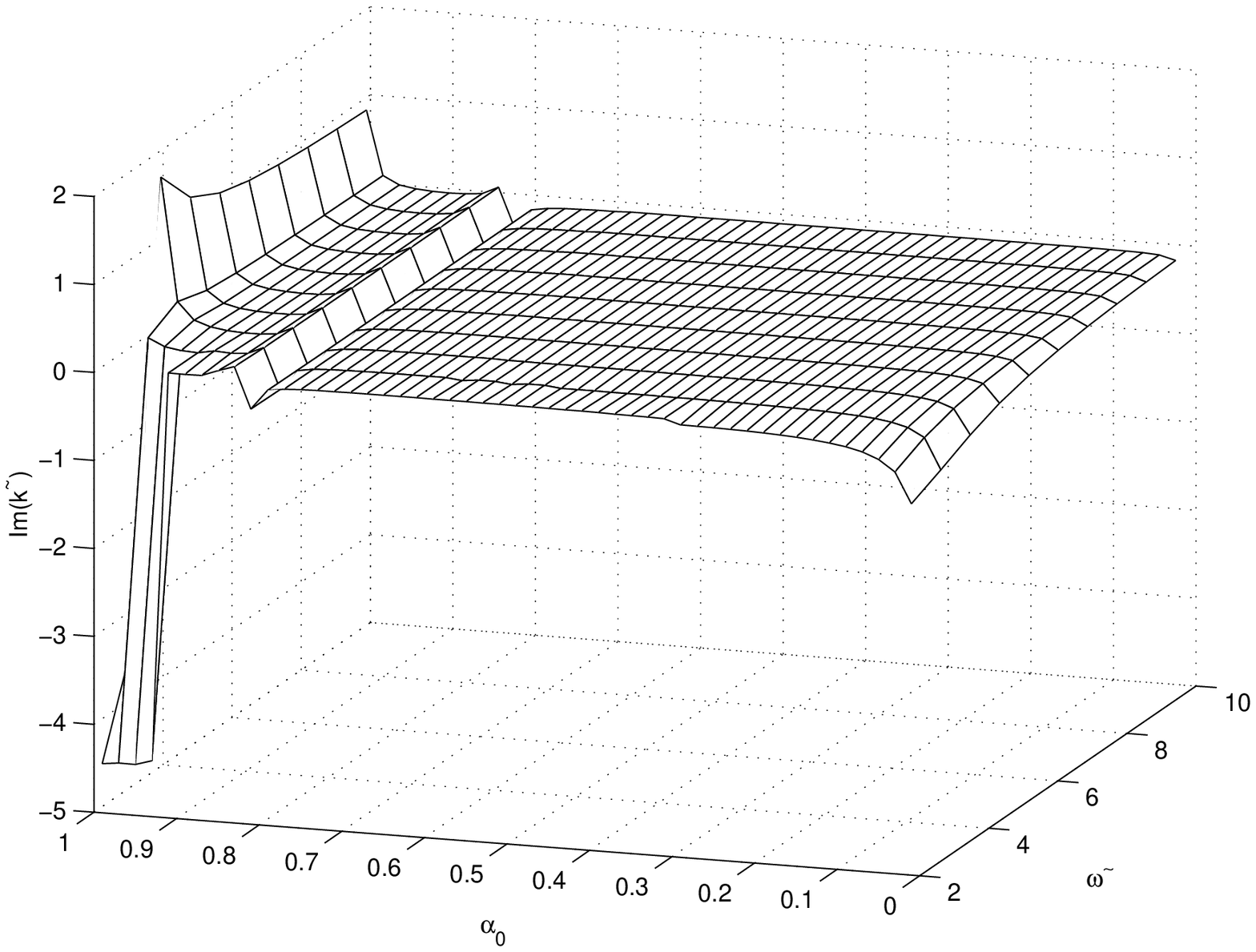}
 \includegraphics[scale=.4]{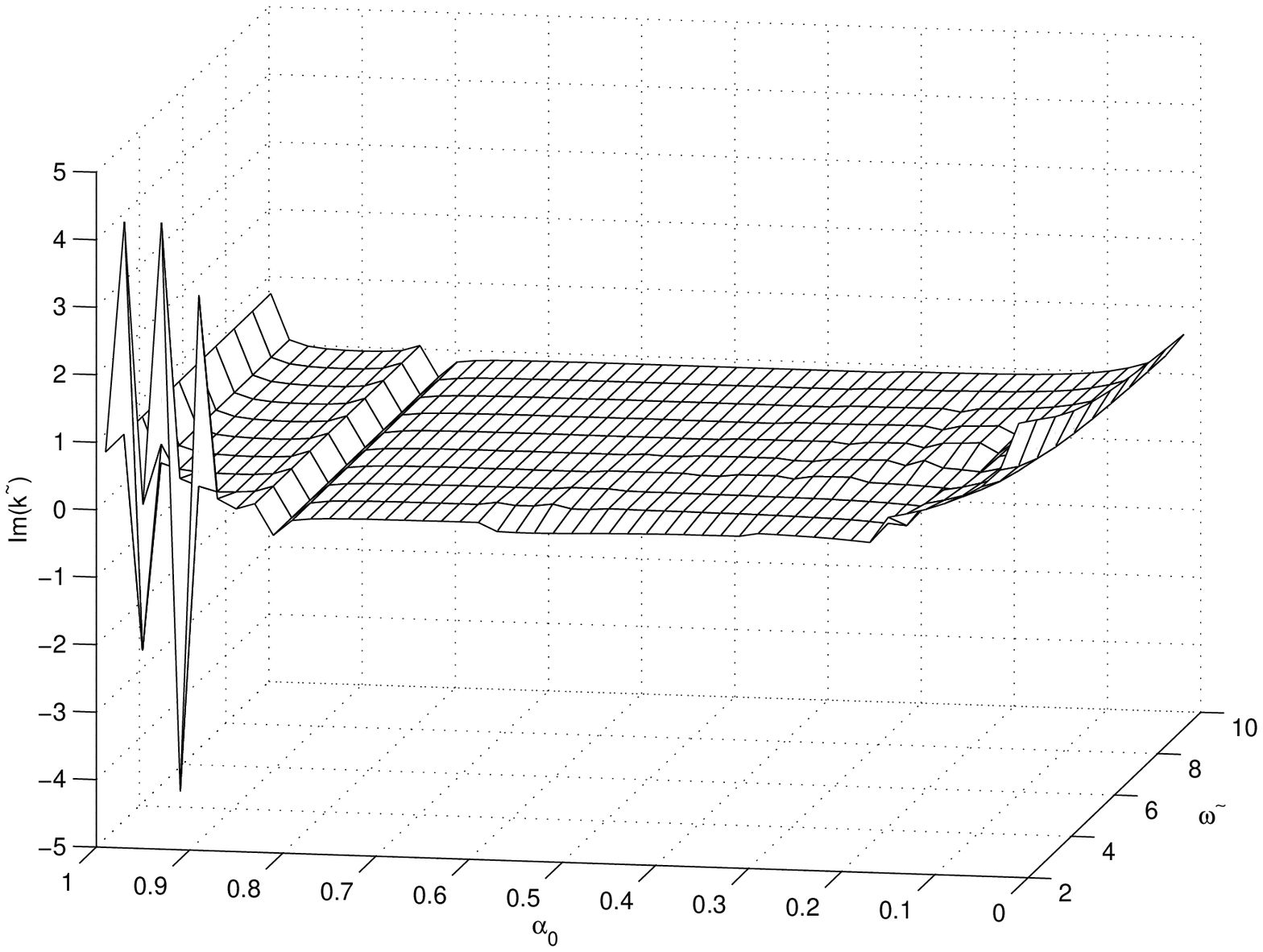}
\end{center}
\caption{\textbf{Top}: Real part for the longitudinal high frequency modes for electron-ion plasma. \textbf{Bottom}: Longitudinal high frequency mode showing both damping and growth with two transonic radii for electron-ion plasma.}\label{fig8}
\end{figure}

\section{Concluding Remarks}\label{sec9}

The prime concern of this study has been exclusively the investigation, within the local approximation, of longitudinal waves in two-fluid plasma situated around the event horizon of the Reissner-Nordstr\"{o}m black hole. The transonic radius begins to play a significant role for the longitudinal waves. From Eqs. (\ref{eq44}) and (\ref{eq45}) it is clear that a singularity take place for each fluid at the point for which the freefall velocity equals the half of the thermal velocity, $u^2_{0s}=\frac{1}{2}v^2_{Ts}$. The position of transonic radius of each fluid mainly dependent on their limiting horizon temperature and determines the temperature at any given radius. Near the transonic radius damping and growth occur for all the low and high frequency modes.

Using a local approximation the dispersion relation for longitudinal waves has been derived. In the limit of zero gravity this result reduced to the special relativistic result obtain by SK \cite{sixteen} for the longitudinal waves.  One interesting point concerning the longitudinal waves in the electron-positron plasma is that, unlike the result found by SK \cite{sixteen} for which only one high frequency mode exist. Here there is no such restriction on the frequency and five low frequency modes are found for both the electron-positron and electron-ion plasmas. Different modes become physical $(Re(k) > 0)$ at the boundary define by the transonic radius of each fluid. It is true for the majority of the modes except for two intriguing low frequency modes for electron-positron plasma that some complicated interplay between frequency and redial distance appears to determine the regions in the $\tilde\omega-\alpha _0 $ plane for which the mode is physical.

The presence of magnetic monopole charge in the RN hole and the characteristic of the extremal RN hole draw attention of the physicists. In view of these reasons, our study of longitudinal waves propagation in a relativistic two-fluid plasma in the environment close to the event horizon of the RN black hole is interesting. The result we obtained reduces to that of the Schwarzschild black hole \cite{fifteen} when $q=0$. Our result can be specialized for the extreme RN hole by choosing $M^2=q^2$. In view of these reasons, our study presented in this paper is thus well motivated.

\noindent
{\large\bf Acknowledgement}\\
The authors (MAR) thanks the Abdus Salam International Centre for Theoretical Physics (ICTP), Trieste, Italy, for giving opportunity to utilize its e-journals for research purpose.

\end{document}